\documentclass[a4paper,11pt]{article}
\pdfoutput=1 
\usepackage{jheppub} 

\usepackage[T1]{fontenc}
\usepackage{slashed}
\usepackage{amssymb}
\usepackage{amsmath}
\usepackage{xcolor}

\title{Off-lightcone Wilson-line operators in gradient flow}

\preprint{TUM-EFT 182/23}

\author[a,b,c]{Nora~Brambilla}
\author[a]{and Xiang-Peng~Wang}

\affiliation[a]{Technical University of Munich, TUM School of Natural Sciences,
Physics Department,\\
James-Franck-Strasse 1, 85748 Garching, Germany} 
\affiliation[b]{Technical University of Munich, Institute for Advanced Study,\\ Lichtenbergstrasse 2a,
85748 Garching, Germany} 
\affiliation[c]{
Technical University of Munich, Munich Data Science Institute, \\ Walther-von-Dyck-Strasse 10, 85748
Garching, Germany}

\emailAdd{nora.brambilla@tum.de}
\emailAdd{xiangpeng.wang@tum.de}

\abstract{Off-lightcone Wilson-line operators are constructed using local operators connected by time-like or space-like Wilson lines, which ensure gauge invariance. Off-lightcone Wilson-line operators have broad applications in various contexts. For instance, space-like Wilson-line operators play a crucial role in determining quasi-distribution functions (quasi-PDFs), while time-like Wilson-line operators are essential for understanding quarkonium decay and production within the potential non-relativistic QCD (pNRQCD) framework.
In this work, we establish a systematic approach for calculating the matching from the gradient-flow scheme to the $\overline{\rm MS}$ scheme in the limit of small flow time for off-lightcone Wilson-line operators. By employing the one-dimensional auxiliary-field formalism, we simplify the matching procedure, reducing it to the matching of local current operators. We provide one-loop level matching coefficients for these local current operators.
For the case of hadronic matrix element related to the quark quasi-PDFs, we show at one-loop level that the finite flow time effect is very small as long as the flow radius is smaller than the physical distance $z$, which is usually satisfied in lattice gradient flow computations. 
Applications include  lattice gradient flow computations of quark/gluon quasi-PDFs,  gluonic correlators related to quarkonium decay and production in pNRQCD, and spin-dependent potentials in terms of chromoelectric and chromomagnetic field insertions into a Wilson loop.}

\begin{document} 
\maketitle
\flushbottom

\section{Introduction}
The path-ordered phase factor $\text{P}\exp\left(ig\int A_{\mu} dx^{\mu}\right)$, referred to as the Wilson line\footnote{Throughout this work, our focus will be exclusively on off-lightcone Wilson lines and Wilson loops, which we will refer to simply as Wilson lines unless ambiguity arises.}, is  a fundamental object  in gauge theories,  playing a crucial role in connecting local fields to construct gauge-invariant operators. The study of the renormalization of  Wilson lines and Wilson loops was initiated by Polyakov \cite{Polyakov:1980ca}  and Gervais, Neveu \cite{Gervais:1979fv}. 
The all-order proof of the multiplicative renormalizability of Wilson lines was made using diagrammatic methods in refs.~\cite{Dotsenko:1979wb,Craigie:1980qs} and functional approach in ref.~\cite{Dorn:1986dt} based on the one-dimensional auxiliary-field formalism \cite{Gervais:1979fv,Arefeva:1980zd, Arefeva:1980uy,Samuel:1978iy,Brandt:1978wc}; see ref.~\cite{Dorn:1986dt} for a review. 

Off-lightcone Wilson-line operators are constructed using local operators connected by time-like or space-like Wilson lines, which insure the gauge invariance of the non-local operators.  Wilson-line operators have broad applications in various contexts. For instance, space-like Wilson-line operators play a crucial role in determining quasi-distribution function (quasi-PDFs) \cite{Ji:2013dva} and pseudo-distribution functions (pseudo-PDFs) \cite{Radyushkin:2017cyf}, while time-like Wilson-line operators are essential for understanding quarkonium decay and production within the potential non-relativistic QCD (pNRQCD) effective theory \cite{Pineda:1997bj,Brambilla:1999xf,Brambilla:2004jw,Brambilla:2001xy,Brambilla:2002nu,Brambilla:2020ojz,Brambilla:2021abf,Brambilla:2022rjd,Brambilla:2022ayc}.

Recently, there has been increasing interests in the study of the quasi-PDFs and pseudo-PDFs defined as the hadronic matrix elements of the Wilson-line operators of the type
\begin{eqnarray}
\label{eq:defqoperator}
\mathcal{O}_{\Gamma}(zv) &=& \bar{\psi}(zv)\Gamma W(zv,0)\psi(0),\\
\label{eq:defgoperator}
\mathcal{O}^{\mu\nu\alpha\beta} (zv)&=& g^2 F^{\mu\nu}(zv)W(zv,0)F^{\alpha\beta}(0),
\end{eqnarray}
where $\psi$ is a quark field, $F^{\mu\nu}$ is the gluon field strength tensor, $\Gamma$ represents a Dirac structure, $v$ is a four-vector satisfying $v^2=-1$ (or $v^2=1$) for space-like (or time-like) Wilson-line operators in Minkowski spacetime \footnote{In this context and throughout subsequent discussions, we stay in Minkowski spacetime. However, a transition to Euclidean spacetime  is warranted when introducing gradient flow, as it is defined in Euclidean spacetime, which indicates $v^2=1$ both for space-like and time-like Wilson-line operators.}, $z$ is a real number and $W(zv, 0)$ is a straight-line path-ordered Wilson line connecting the two fields,
\begin{eqnarray}
W(zv, 0) = \text{P}\exp\left(ig\int_0^z ds\, v\cdot A(sv) \right),
\end{eqnarray}
in which the fundamental representation and the adjoint representation of the gauge group are assumed for the quark case, and the gluon case, respectively. Without loss of generality, we choose $z>0$ throughout this work. Notably,  quasi-PDFs and pseudo-PDFs can be factorized in relation to the standard light-cone parton distribution functions (PDFs) based on the short-distance operator product expansion (OPE) and the large momentum effective theory (LaMET) \cite{Braun:2007wv, Ji:2014gla,Izubuchi:2018srq}.  Employing hadronic matrix elements of  off-lightcone Wilson-line operators as opposed to lightlike ones, offers the advantage of being considerably more suitable to lattice computations. However, this convenience comes with the trade-off of the highly non-trivial renormalization of the off-lightcone Wilson-line operators. For a comprehensive overview, we refer to ref.~\cite{Ji:2020ect}.
Furthermore, the vacuum expectation value of gluonic Wilson-line operators, such as the one defined in eq.~(\ref{eq:defgoperator}), are related to the studies of quarkonium decay and production in the framework of pNRQCD, and the QCD vacuum structure 
(see ref.~\cite{DiGiacomo:2000irz} for a review).

The gradient flow was originally introduced as  a method to regularize and suppress the ultraviolet (UV) fluctuations \cite{Narayanan:2006rf,Luscher:2009eq} in lattice calculations. It has proven useful in lattice QCD scale settings and in lattice calculation of local matrix elements and correlation functions \cite{Luscher:2010iy,Luscher:2011bx,Luscher:2013cpa,Luscher:2013vga,Borsanyi:2012zs,Sommer:2014mea,Suzuki:2013gza,Makino:2014taa,Harlander:2018zpi,FlavourLatticeAveragingGroupFLAG:2021npn,Leino:2021vop,Mayer-Steudte:2022uih,Leino:2022kgj}. Alongside its noise-reducing effect, gradient flow exhibits the notable property that composite operators at positive flow time remain finite even in the continue limit,  provided they are expressed using renormalized physical parameters and fields \cite{Luscher:2011bx}. This property renders gradient flow suitable as a matching scheme between lattice and perturbative calculations. It is worth noting that gradient flow does not modify the infrared behavior of operators in the small flow-time limit, allowing the use of dimensional regularization to regulate infrared (IR) divergences when matching from the gradient-flow scheme to $\overline{\rm MS}$-like schemes.

The initial proposal to explore quasi-PDFs with the incorporation of gradient flow was presented in ref.~\cite{Monahan:2016bvm}. Subsequently, a one-loop matching calculation was performed, considering full flow time dependencies, for the hadronic matrix element defined using the operator in eq.~(\ref{eq:defqoperator}) with zero external momenta, which is related to the quark quasi-PDF \cite{Monahan:2017hpu}. In this study, we present a systematic approach for calculating the matching from the gradient-flow scheme to the $\overline{\rm MS}$ scheme for Wilson-line operators, such as those defined in eq.~(\ref{eq:defqoperator}) and eq.~(\ref{eq:defgoperator}), specifically in the limit of small flow time. Our approach is based on the one-dimensional auxiliary-field formalism, and we provide comprehensive details of the one-loop order matching calculations.
For more recent applications of perturbative gradient flow, see refs.~\cite{Artz:2019bpr,Rizik:2020naq,Suzuki:2020zue,Harlander:2020duo,Boers:2020uqr,Shindler:2021bcx,Harlander:2022tgk,Mereghetti:2021nkt,Harlander:2022vgf,Brambilla:2021egm,Crosas:2023anw} and the references therein.

We structure the remaining content of the paper as follows. In section \ref{sec:EFT}, we introduce the one-dimensional auxiliary-field formalism and elaborate on its application in the renormalization of Wilson-line operators. In section \ref{sec:introGF}, we introduce the gradient-flow formalism and implement the small flow-time expansion to the local current operators, yielding the core formulas for our matching calculations. Section \ref{sec:oneloopcalc} is dedicated to the computation of the matching coefficients up to one-loop order. Our summary and conclusions can be found in section \ref{sec:summary}. In appendix \ref{app:quarkquasi}, we present the detailed matching calculation with full flow time dependence for the hadronic matrix element that is related to quasi-PDFs at zero external momenta. We compare these findings with existing results from ref.~\cite{Monahan:2017hpu}, as well as with the results of section \ref{sec:oneloopcalc} in the small flow-time limit.

\section{One-dimensional auxiliary-field formalism and renormalization of the off-lightcone Wilson-line operators}
\label{sec:EFT}
\subsection{The one-dimensional auxiliary-field formalism}
The one-dimensional auxiliary-field formalism is defined by enlarging the QCD Lagrangian to include an extra term \cite{Dorn:1986dt},
\begin{eqnarray}
\label{eq:extendLBare}
\mathcal{L}_{h_v} =\bar{h}_{v, 0} (iv\cdot D) h_{v, 0}, 
\end{eqnarray}
where the subscript ``$0$”  indicates a bare quantity, $h_{v, 0}$ is an auxiliary ``heavy”  Grassmann (or complex) scalar field in either fundamental or adjoint representations of the SU($3$) gauge group, $D_0^{\mu} = \partial^{\mu} - ig_{0}A_0^{\mu, a}T^a$ is the  covariant derivative and $T^a$ are the SU($3$) generators in the appropriate representation. Note that, for $v=(1,0,0,0)$ and $h_{v, 0}$ being in the fundamental representation, the effective Lagrangian in eq.~(\ref{eq:extendLBare}) essentially coincides with the Lagrangian of the heavy-quark effective theory (HQET) in the rest frame \cite{Eichten:1989zv}, except that we choose $h_{v, 0}$ to be a  scalar.

The effective Lagrangian $\mathcal{L}_{h_v}$ is amenable to renormalization through the redefinitions,
\begin{eqnarray}
h_{v, 0} = Z_{h_v}^{\frac{1}{2}} h_v,\, \, \, \, g_0 =Z_g g,\, \, \, \,   A_0 = Z_A^{\frac{1}{2}} A,
\end{eqnarray}
where  $Z_{h_v}$ is the wave-function renormalization constant of the ``heavy” auxiliary field, $Z_g$ and $Z_A$ are the usual QCD renormalization constants of the strong coupling and the gluon field, respectively. There will be a ``mass correction” $i\delta m$ for the ``heavy” auxiliary field and the effective Lagrangian $\mathcal{L}_{h_v} $ expressed in terms of  the renormalized fields and parameters is given by \footnote{The ``mass correction” $i\delta m$ is imaginary for space-like $v$, but real for time-like $v$ in Minkowski spacetime. In Euclidean spacetime,  $i\delta m$ is alway  imaginary. },
\begin{eqnarray}
\label{eq:extendLRen}
\mathcal{L}_{h_v} =Z_{h_v}\bar{h}_v (iv\cdot \partial - i\delta m) h_v + gZ_gZ_A^{\frac{1}{2}}Z_{h_v} \bar{h}_v v\cdot A^{a}T^a h_v,
\end{eqnarray}
where proper representations are implied for the fields and the generator $T^a$. 

The ``mass correction” $i\delta m$ is linearly divergent. The linear divergence vanishes in dimensional regularization, but persists, in lattice regularization and the gradient-flow formalism. Furthermore, the ``mass correction” $i\delta m$ could incorporate contributions of $\mathcal{O}(\Lambda_{\rm QCD})$, attributed to the renormalon ambiguities that are recognized to be present in HQET \cite{Bigi:1994em, Beneke:1994sw}. 

Based on the one-dimensional auxiliary-field formalism, up to a constant normalization factor, the connected two-point function of the ``heavy” auxiliary fields can be related to the Wilson lines through \cite{Dorn:1986dt,Braun:2020ymy,Ji:2020ect},
\begin{eqnarray}
\langle h_{v,0}(x)\bar{h}_{v,0}(0) \rangle_{h_v} = W\left(\frac{v\cdot x}{v^2}, 0\right)\theta\left(\frac{v\cdot x}{v^2}\right)\delta^{(d-1)}(x_\perp),
\end{eqnarray}
where $\langle ... \rangle_{h_v}$ stands for integrating out the ``heavy” auxiliary field $h_v$,  and $x_{\perp}^{\mu} = x^{\mu} - v^{\mu}(v\cdot x)/v^2$.
In this way,  the Wilson-line operators can be substituted with products of local current operators. For the Wilson-line operators defined in eq.~(\ref{eq:defqoperator}) and eq.~(\ref{eq:defgoperator}), we have
\begin{eqnarray}
\mathcal{O}^{\rm B}_{\Gamma}(zv)  &=& \int d^d x\, \delta\left(\frac{v\cdot x}{v^2} -z\right)\langle \bar{\psi}_{0}(x)h_{v,0}(x) \Gamma \bar{h}_{v,0}(0)\psi_0(0)\rangle_{h_v},\\
\mathcal{O}^{\mu\nu\alpha\beta, \rm B}(zv)  &=& \int d^d x\, \delta\left(\frac{v\cdot x}{v^2} -z\right)\langle g^2F_{0}^{\mu\nu}(x)h_{v,0}(x)  \bar{h}_{v,0}(0)F_{0}^{\alpha\beta}(0)\rangle_{h_v},
\end{eqnarray}
where the superscript ``B” indicates bare composite operators, the bare ``heavy” auxiliary fields $h_{v,0}$ in $\mathcal{O}^{\rm B}_{\Gamma}(zv) $, $\mathcal{O}^{\mu\nu\alpha\beta, \rm B}(zv)$ are in fundamental representation and adjoint representation, respectively.

\subsection{Renormalization of the off-lightcone Wilson-line operators}
The operator as defined in eq.~(\ref{eq:defgoperator}) is not renormalized multiplicatively because, in the presence of the external four-vector $v^{\mu}$,  the components parallel and transverse to $v^{\mu}$ are renormalized differently \cite{Braun:2020ymy}. We introduce the following projectors,
\begin{eqnarray}
g^{\mu\nu}_{\|} = \frac{v_{\mu}v_{\nu}}{v^2},\, \, \, \, g^{\mu\nu}_{\perp} = g^{\mu\nu} - \frac{v_{\mu}v_{\nu}}{v^2},
\end{eqnarray}
to project out the parallel and the transverse components of the field strength tensor $F^{\mu\nu}$ by defining,
\begin{eqnarray}
F^{\mu\nu}_{\|\perp} &=& g^{\mu\alpha}_{\|}g^{\nu\beta}_{\perp}F^{\alpha\beta},\\
F^{\mu\nu}_{\perp\perp} &=& g^{\mu\alpha}_{\perp}g^{\nu\beta}_{\perp}F^{\alpha\beta}.
\end{eqnarray}
Notice that we ignore the mixing with gauge-noninvariant operators since they do not contribute to gauge-invariant observables. Nevertheless, we will verify the mixing patterns found in refs.~\cite{Dorn:1981wa,Zhang:2018diq,Wang:2019tgg} whenever they are relevant for our computations.
Obviously, when $v=(1,0,0,0)$ and $\mu\neq\nu$, $F^{\mu\nu}_{\|\perp}, F^{\mu\nu}_{\perp\perp}$ are proportional to the chromoelectric field \textbf{E} and the chromomagnetic field \textbf{B}, respectively. We introduce the following gluonic Wilson-line operators,
\begin{eqnarray}
\label{eq:defpperp}
\mathcal{O}_{\|\perp}^{\mu\nu\alpha\beta} (zv)&=& g^2 F_{\|\perp}^{\mu\nu}(zv)W(zv,0)F_{\|\perp}^{\alpha\beta}(0),\\
\label{eq:defperpperp}
\mathcal{O}_{\perp\perp}^{\mu\nu\alpha\beta} (zv)&=& g^2 F_{\perp\perp}^{\mu\nu}(zv)W(zv,0)F_{\perp\perp}^{\alpha\beta}(0).
\end{eqnarray}
The off-lightcone Wilson-line operators defined in eq.~(\ref{eq:defqoperator}), eq.~(\ref{eq:defpperp}), and eq.~(\ref{eq:defperpperp})  undergo multiplicative renormalization within the auxiliary-field formalism, as proved in refs.~\cite{Green:2017xeu, Ji:2017oey, Zhang:2018diq}, and shown diagrammatically in refs.~\cite{Ishikawa:2017faj, Li:2018tpe}. That is, the renormalized off-lightcone Wilson-line operators in eq.~(\ref{eq:defqoperator}), eq.~(\ref{eq:defpperp}) and eq.~(\ref{eq:defperpperp}) are related to the bare operators by the formulas,
\begin{eqnarray}
\mathcal{O}^{\rm B}_{\Gamma}(zv, \Lambda) &=& Z_{q}(\Lambda,\mu) e^{\delta m(\Lambda)z} \mathcal{O}^{\rm R}_{\Gamma}(zv, \mu),\\
\mathcal{O}^{\mu\nu\alpha\beta,\, \rm B}_{\|\perp}(zv, \Lambda) &=& Z_{\|\perp}(\Lambda,\mu) e^{\delta m(\Lambda)z} \mathcal{O}_{\|\perp}^{\mu\nu\alpha\beta,\, \rm R}(zv, \mu),\\
\mathcal{O}^{\mu\nu\alpha\beta,\, \rm B}_{\perp\perp}(zv, \Lambda) &=& Z_{\perp\perp}(\Lambda,\mu) e^{\delta m(\Lambda)z} \mathcal{O}_{\|\perp}^{\mu\nu\alpha\beta,\, \rm R}(zv, \mu),
\end{eqnarray}
where the superscript ``R” stands for renormalized composite operators, $\mu$ is the renormalization scale, $\Lambda$ represents a UV cutoff (UV regulator), $\delta m(\Lambda)$ encodes the ``mass correction” of the Wilson line, with its  its  linear divergence, while $Z_{q}(\Lambda,\mu)$, $Z_{\|\perp}(\Lambda,\mu)$ and $Z_{\perp\perp}(\Lambda,\mu)$ encode all logarithmic divergences originating from wave-function and vertex renormalizations.

Within the one-dimensional auxiliary-field formalism, after subtracting linear divergences, the renormalization of the off-lightcone Wilson-line operators defined in eq.~(\ref{eq:defqoperator}), eq.~(\ref{eq:defpperp}), and eq.~(\ref{eq:defperpperp}) is simplified into the renormalization of two local ``heavy-to-light” or ``heavy-to-gluon” currents \footnote{See refs.~\cite{Stefanis:1983ke,Stefanis:2012if} for the renormalization of off-lightcone Wilson-line operators in a different approach based on the Mandelstam formalism.},
\begin{eqnarray}
\label{eq:currentoperator}
J_q(x) &=& \bar{\psi}(x) h_v(x),\\
J^{\mu\nu}_{\|\perp}(x) &=& gF_{\|\perp}^{\mu\nu}(x)h_v(x),\\
J^{\mu\nu}_{\perp\perp}(x) &=& gF_{\perp\perp}^{\mu\nu}(x)h_v(x).
\end{eqnarray}
The renormalization formulas for the above  gauge-invariant local current operators are given by,
\begin{eqnarray}
J_q^{\rm B} (x, \Lambda) &=& Z_{J_q} (\Lambda,\mu)J_q^{\rm R}(x, \mu),\\
J^{\mu\nu,\, \rm B}_{\|\perp}(x, \Lambda) &=& Z_{J,\, \|\perp}(\Lambda,\mu)J^{\mu\nu,\, \rm R}_{\|\perp}(x, \mu),\\
J^{\mu\nu,\, \rm B}_{\perp\perp}(x, \Lambda) &=& Z_{J,\, \perp\perp}(\Lambda,\mu)J^{\mu\nu,\, \rm R}_{\perp\perp}(x, \mu),
\end{eqnarray}
leading to the equivalence 
\begin{eqnarray}
Z_{q} (\Lambda,\mu)&=& Z_{J_q}^2(\Lambda,\mu),\\
Z_{\|\perp} (\Lambda,\mu)&=& Z^2_{J,\, \|\perp}(\Lambda,\mu),\\
Z_{\perp\perp} (\Lambda,\mu)&=& Z^2_{J,\, \perp\perp}(\Lambda,\mu).
\end{eqnarray}
In the following text, we may omit indicating the arguments $\Lambda$ and $\mu$ in the renormalization constants.

The renormalization constants for the local current operators can be further decomposed as
\begin{eqnarray}
Z_{J_q} &=& Z_{\psi}^{\frac{1}{2}} Z_{h_v}^{\frac{1}{2}}Z^V_{J},\\
Z_{J,\, \|\perp} &=& Z_{A}^{\frac{1}{2}} Z_{h_v}^{\frac{1}{2}}Z^V_{\|\perp},\\
Z_{J,\, \perp\perp} &= & Z_{A}^{\frac{1}{2}} Z_{h_v}^{\frac{1}{2}}Z^V_{\perp\perp},
\end{eqnarray}
where $Z^V_{J}$, $Z^V_{|\perp}$, and $Z^V_{\perp\perp}$ represent the corresponding renormalization factors arising from vertex corrections. Implicit in this notation is the dependence of the wave-function renormalization constant of the  ``heavy” auxiliary field from the color representations. 

The Wilson-line operators that define the spin-dependent potentials in terms of chromomagnetic field insertions into a Wilson loop \cite{Brambilla:2000gk,Pineda:2000sz} are also of interest in this study. Within the auxiliary-field formalism, each insertion of a chromomagnetic field into a Wilson loop can be related to the following current operator as, 
\begin{eqnarray}
\label{eq:currentoperatorspin}
J^{\mu\nu}_{F}(x) &=&\bar{h}_v(x)gF^{\mu\nu}_{\perp\perp}(x)h_v(x),
\end{eqnarray}
where $v=(1, 0,0,0)$ and the subscript $F$ indicates that the ``heavy” auxiliary fields are in the fundamental representation of the SU($3$)  group. The renormalization formula for this operator is , 
\begin{eqnarray}
J^{\mu\nu,\, \rm B}_{F}(x)&=&Z_{F}J^{\mu\nu,\, \rm R}_{F}(x) = Z_{A}^{1/2} Z_{h_v}Z^V_{F}J^{\mu\nu,\, \rm R}_{F}(x),
\end{eqnarray}
where $Z^V_{F}$ represents the renormalization factor stemming from the vertex corrections of the current $\bar{h}_v(x)gF^{\mu\nu}_{\perp\perp}(x)h_v(x)$.

\section{Gradient flow and small flow-time expansion}
\label{sec:introGF}
\subsection{The gradient-flow formalism}
In the following, we work in $d$-dimensional Euclidean spacetime with $d=4-2\epsilon$.
The gradient flow extends the $d$-dimensional QCD to a $(d+1)$-dimensional field theory by introducing an additional coordinate $t$ (not to be confused with the time coordinate in Minkowski spacetime) of mass dimension $-2$,  which is called the flow time \cite{Luscher:2010iy, Luscher:2011bx}. In the gradient-flow formalism, the flow time dependent gluon field $B_{\mu}^a(x, t)$ and quark field $\chi^i_{\alpha}(x, t)$ 
are determined by the flow equations \cite{Luscher:2010iy, Luscher:2011bx,Luscher:2013cpa},
\begin{eqnarray}
\label{eq:flowequations}
\partial_t B_{\mu} &=& \mathcal{D}_{\nu}G_{\nu\mu} +\kappa \mathcal{D}_{\mu}\partial_{\nu}B_{\nu},\\
\partial\chi &= & \mathcal{D}_{\mu}\mathcal{D}_{\mu}\chi -\kappa (\partial_{\mu}B_{\mu})\chi,\\
\partial\bar{\chi} &=& \bar{\chi}\overleftarrow{\mathcal{D}}_{\mu}\overleftarrow{\mathcal{D}}_{\mu} +\kappa\bar{\chi}\partial_{\mu}B_{\mu},
\end{eqnarray}
where  $\kappa$ is a gauge parameter that drops out in physical observables, and
\begin{eqnarray}
G_{\mu\nu} &= &\partial_{\mu}B_{\nu} -\partial_{\nu}B_{\mu} + \left[B_{\mu}, B_{\nu}\right],\\
\mathcal{D}_{\mu} &=&\partial_{\mu} + \left[B_{\mu}, .\right],
\end{eqnarray}
with the boundary conditions
\begin{eqnarray}
\label{eq:boundary}
B_{\mu}(x; t=0) &=& gA_{\mu}(x),\\
\chi(x; t=0) &=& \psi(x),\\
\bar{\chi}(x; t=0) &=& \bar{\psi}(x).
\end{eqnarray}

In perturbative gradient flow, one extends the regular QCD Feynman rules by adding flow time dependent exponentials to the propagators, introducing  extra ``flow lines” and ``flow vertices” \cite{Luscher:2011bx}; see ref.~\cite{Artz:2019bpr} for more details about the gradient flow Feynman rules and higher order calculation techniques. In the following perturbative calculation, we  adopt the  convenient choice $\kappa=1$.

With renormalized physical parameters, the flowed field $B_{\mu}$ does not need further renormalization \cite{Luscher:2010iy,Luscher:2011bx}, but the flowed quark fields $\chi$ need to be  renormalized \cite{Luscher:2013cpa},
\begin{eqnarray}
\chi_{0}(x; t) = Z_{\chi}^{\frac{1}{2}} \chi(x; t), \, \, \bar{\chi}_0(x; t) = Z_{\chi}^{\frac{1}{2}} \bar{\chi}(x; t), \, \,  Z_{\chi} = 1- \frac{\alpha_s}{4\pi}C_F \frac{3}{\epsilon_{\rm UV}} + O(\alpha_s^2),
\end{eqnarray}
where $C_F = (N_c^2-1)/(2N_c)$ with $N_c=3$ the number of colors.
In order to avoid the complication due to the matching between $ Z_{\chi}$ in dimensional regularization and that in the lattice regularization, the ringed fermion fields $\mathring{\chi}(x,t)$, $\mathring{\bar{\chi}}(x,t)$ were introduced in ref.~\cite{Makino:2014taa},
\begin{eqnarray}
\mathring{\chi}(x,t) = \mathring{Z}_{\chi}^{\frac{1}{2}}\chi_{0}(x,t),\, \,  \, \mathring{\bar{\chi}}(x,t) = \mathring{Z}_{\chi}^{\frac{1}{2}}\bar{\chi}_{0}(x,t),
\end{eqnarray}
with
\begin{eqnarray}
 \mathring{Z}_{\chi}=\frac{-2N_c}{(4\pi)^2t^2
\langle \bar{\chi}_{0}(x,t)\overleftrightarrow{\slashed{\mathcal{D}}}\chi_{0}(x,t)\rangle|_{m=0}},
\end{eqnarray}
for one quark flavor, where $\overleftrightarrow{\mathcal{D}}^{\mu} = {\mathcal{D}}^{\mu}  - \overleftarrow{\mathcal{D}}^{\mu}$. The next-next-to-leading order (nnLO) result of $\mathring{Z}_{\chi}$ can be found in  ref.~\cite{Artz:2019bpr}. At next-to-leading order (NLO), the renormalization factor $ \mathring{Z}_{\chi}$ is given by \cite{Makino:2014taa}
\begin{eqnarray}
\label{eq:ringedfermionren}
\mathring{Z}_{\chi}(t, \mu) = (8\pi t)^{-\epsilon}\left\{1+  \frac{\alpha_s}{4\pi}C_F\left[\frac{3}{\epsilon_{\rm UV}} + 3\log\left(2\mu^2 te^{\gamma_E}\right)-\log(432)\right] \right\},
\end{eqnarray}
where  the factor $(8\pi t)^{-\epsilon}$ is introduced to balance the dimension of the ringed fermion field, which has dimension $3/2$, while the flowed fermion field has dimension $(d-1)/2$.

\subsection{The small flow-time expansion}
\label{subsec:smallflowexp}
In this work, we consider the local current operators defined in eqs.~(\ref{eq:currentoperator}) and eq.~(\ref{eq:currentoperatorspin}), which are flowed at the flow time $t$. We use the same symbols for both the flowed and the un-flowed current operators as long as it does not lead to confusions. In the context of Wilson-line operators, we flow the external fields and the Wilson lines at the flow time $t$. Specifically, this means that all the gluon fields directly connecting to the ``heavy” auxiliary field lines are flowed at the flow time $t$, while the ``heavy” auxiliary fields themselves remain un-flowed. In the regime of small flow time, we employ a ``short-distance”  OPE  for the flowed current operators. Our objective is to derive matching coefficients between renormalized flowed current operators and the  $\overline{\rm MS}$ renormalized un-flowed current operators,
\begin{eqnarray}
\label{eq:smalltexpansion}
\mathcal{O}^{\rm R} (t)= c_{\mathcal{O}}(t, \mu)\mathcal{O}^{\overline{\rm MS}} (\mu) + O(t),
\end{eqnarray}
where we have neglected operator mixing.

We work in the massless limit for quark fields and set all the external scales to be zero whenever feasible. 
This strategic choice allows us to leverage established techniques from HQET for matching calculations.  Specifically, we do not need to calculate the un-flowed operators because they yield scaleless integrals to all orders in perturbative calculations in dimensional regularization so that we can obtain the matching coefficients $c_{\mathcal{O}}(t, \mu)$ from the renormalized flowed operators simply by subtracting the IR poles.

Similarly to the renormalization of current operators, we can break down the matching coefficients $c_{\mathcal{O}}(t, \mu)$ for the local current operators into products of field matching coefficients and corresponding vertex matching coefficients. For the four local current operators $\bar{\psi} h_v$, $gF^{\mu\nu}_{\|\perp}h_v$, $gF^{\mu\nu}_{\perp\perp}h_v$, and $g\bar{h}_vF^{\mu\nu}_{\perp\perp}h_v$, the relationships between matching coefficients for the local current operators and those for the corresponding fields and vertices are given by
\begin{subequations}
\label{eq:operatorrelations}
\begin{eqnarray}
c_{\psi h_v} (t,\mu) &=& \mathring{\zeta}_{\psi}(t,\mu)\zeta_{h_v}^F(t,\mu)\zeta_{\psi h_v}(t,\mu),\\
c_{\|\perp} (t,\mu)&=& \zeta_{A}(t,\mu)\zeta_{h_v}^A(t,\mu)\zeta_{\|\perp}(t,\mu),\\
c_{\perp\perp}(t,\mu) &=& \zeta_{A}(t,\mu)\zeta_{h_v}^A(t,\mu)\zeta_{\perp\perp}(t,\mu),\\
c_F(t,\mu) &=& \zeta_{A}(t,\mu)(\zeta^F_{h_v}(t,\mu))^2\zeta_F(t,\mu).
\end{eqnarray}
\end{subequations}
Here $ \mathring{\zeta}_{\psi}(t,\mu)$, $\zeta_{h_v}^F (\zeta_{h_v}^A)$ and $\zeta_{A}(t,\mu)$ are the matching coefficients of the ringed fermion field $\mathring{\chi}$,  the ``heavy” auxiliary field $h_v$ in fundamental (adjoint) representation and the flowed gluon field $B$, respectively. Additionally, $\zeta_{\psi h_v}(t,\mu)$, $\zeta_{\|\perp}(t,\mu)$, $\zeta_{\perp\perp}(t,\mu)$ and $\zeta_F(t,\mu)$ denote matching coefficients stemming from vertex corrections for the corresponding flowed local current operators. We calculate these matching coefficients to one-loop order in section \ref{sec:oneloop}.

A key insight arising from our study in section \ref{sec:EFT} on the multiplicative renormalization of Wilson-line operators is that matching coefficients for Wilson-line operators in the transition from the gradient-flow scheme to the $\overline{\rm MS}$ scheme can be directly derived from matching coefficients for the corresponding local current operators. This occurs because in the one-dimensional auxiliary-field formalism, combinations of the renormalized local current operators at different spacetime points do not introduce additional UV divergences, apart from the subtracted linear divergence, and because of eq.~(\ref{eq:smalltexpansion}) that relates local operators in the small flow-time limit. Therefore, additional matching for these combined local current operators in different spacetime points is not needed in the small flow-time limit.

\section{Computations and results}
\label{sec:oneloopcalc}
\label{sec:oneloop}
While the matching coefficients for the fields can be gauge-dependent, it is important to note that the combinations leading to the local current operators under consideration in this work are gauge invariant. Therefore, we have chosen to employ the Feynman gauge consistently throughout our calculations. Our computations are conducted in the context of Euclidean spacetime. In Euclidean spacetime, the renormalized Lagrangian $\mathcal{L}_{h_v}$ is given by
\begin{eqnarray}
\label{eq:extendLRenEuclidean}
\mathcal{L}_{h_v}^{\rm E} =Z_{h_v}\bar{h}_v (-iv\cdot \partial + i\delta m) h_v + gZ_gZ_A^{\frac{1}{2}}Z_{h_v} \bar{h}_v v\cdot A^{a}T^a h_v,
\end{eqnarray}
where the superscript ``E” indicates Euclidean spacetime. Consequently, 
the Feynman rules for the ``heavy” auxiliary field propagator, with momentum $k$ flowing along the ``heavy” auxiliary field line,  and the  interaction vertex are given by 
$\frac{1}{k\cdot v -i0}$ and $-gv^{\mu}T^a$, respectively. The $-i0$ prescription in Euclidean spacetime plays key role in obtaining correct linear divergences and ``mass correction”  $i\delta m$.

We decompose a vector into parallel and transverse components according to ($v^2=1$ in Euclidean spacetime in our settings)
\begin{eqnarray}
q_{\|}^{\mu} = (v\cdot q) v^{\mu}, \, \, \, \, q_{\perp}^{\mu} = q^{\mu} - v\cdot q v^{\mu}.
\end{eqnarray}
We use the following Schwinger parametrization ($\alpha$-representation)
\begin{eqnarray}
\label{eq:alphax}
\frac{1}{A^n} = \frac{1}{\Gamma(n)}\int_0^{+\infty} d\alpha\, \alpha^{n-1} e^{-A\alpha},
\end{eqnarray}
for the QCD propagator denominators. For the eikonal propagator, we use another version of Schwinger parametrization
\begin{eqnarray}
\label{eq:alphay}
\frac{1}{(k\cdot v- i0)^n} = \frac{(i)^n}{\Gamma(n)}\int_0^{+\infty} d\alpha\, \alpha^{n-1} e^{-i(k\cdot v)\alpha},
\end{eqnarray}
with $k\cdot v$  real.

\subsection{$\delta m$ \& $\zeta_{h_v}$}
\begin{figure}
\centering
\includegraphics[width =0.32\textwidth]{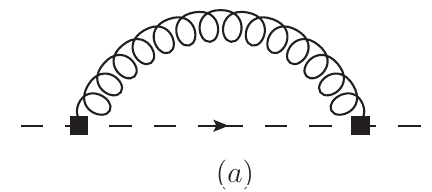}
\caption{One-loop self-energy Feynman diagram for the ``heavy” auxiliary field in gradient-flow formalism. The dashed arrow line denotes the ``heavy” auxiliary field, the filled squares represent the flowed $B_{\mu}$ field at flow time $t$.}
\label{fig:heavyself}
\end{figure}
The calculation of the self-energy of the ``heavy” auxiliary field at one-loop in gradient flow is similar to that of the heavy quark field in HQET.
Setting the external momentum $p$ flowing from left to right, the one-loop self-energy diagram shown in figure \ref{fig:heavyself} gives the amplitude
\begin{eqnarray}
\mathcal{M}_{(a)} = g^2\tilde{\mu}^{2\epsilon}C_F\int\frac{d^dk}{(2\pi)^d}\frac{1}{\left(k\cdot v + p\cdot v - i0\right) k^2 } e^{-2tk^2},
\end{eqnarray}
where $\tilde{\mu}^{2\epsilon}= \left(\frac{\mu^2 e^{\gamma_E}}{4\pi} \right)^{\epsilon}$ and we have chosen  the ``heavy” auxiliary field to be in fundamental representation. For the adjoint representation case, we just need to replace $C_F$ with $C_A$ in the above expression.

Summing up geometric series of the one-loop self-energy diagrams of the ``heavy” auxiliary field propagator gives $1/(p\cdot v - \mathcal{M}_{(a)})$, which indicates that the one-loop ``mass correction” is given by $(-\mathcal{M}_{(a)})$. For simplicity, we set $p=0$ in $\mathcal{M}_{(a)}$, yielding
\begin{eqnarray}
\label{eq:deltamstart}
-\mathcal{M}_{(a)}|_{p=0}= -g^2\tilde{\mu}^{2\epsilon}C_F\int\frac{d^dk}{(2\pi)^d}\frac{1}{\left(k\cdot v- i0\right)k^2} e^{-2tk^2}.
\end{eqnarray}
Note that the above expression is not odd under $k\cdot v\rightarrow -k\cdot v$ and the $k\cdot v$ integration does not give a vanishing result  because the sign of the $i0$ prescription matters. 

 To derive the matching coefficient of the  ``heavy” auxiliary field at one-loop, we differentiate $\mathcal{M}_{(a)} $ with respect to $p\cdot v$,  and set $p=0$ for simplicity, resulting in
\begin{eqnarray}
\label{eq:Zhvstart}
\frac{\partial\mathcal{M}_{(a)}}{\partial (p\cdot v)}|_{p=0}= - g^2\tilde{\mu}^{2\epsilon}C_F\int\frac{d^dk}{(2\pi)^d}\frac{1}{\left(k\cdot v- i0\right)^2k^2} e^{-2tk^2}.
\end{eqnarray}

Now, we explicitly evaluate eq.~(\ref{eq:deltamstart}) and eq.~(\ref{eq:Zhvstart}) using the Schwinger parameterization in eq.~(\ref{eq:alphax}) and eq.~(\ref{eq:alphay}). We find (we set $\epsilon=0$ whenever the integral is finite)
\begin{eqnarray}
\label{eq:deltamcalc0}
-\mathcal{M}_{(a)}|_{p=0} &=& -ig^2C_F\int\frac{d^4k}{(2\pi)^4}\int_0^{+\infty}dx\int_0^{+\infty}dy\, e^{-(2t+x)k^2- i(k\cdot v)y},\\
\label{eq:Zhvcalc0}
\frac{\partial\mathcal{M}_{(a)}}{\partial (p\cdot v)}|_{p=0} &=& g^2\tilde{\mu}^{2\epsilon}C_F\int\frac{d^dk}{(2\pi)^d}\int_0^{+\infty}dx\int_0^{+\infty}dy\, y\, e^{-(2t+x)k^2- i(k\cdot v)y},
\end{eqnarray}
Performing the $k$ momentum integration gives 
\begin{eqnarray}
\label{eq:deltamcalc1}
-\mathcal{M}_{(a)}|_{p=0} &=& -i\frac{\alpha_s}{4\pi}C_F\int_0^{+\infty}dx\int_0^{+\infty}dy\, \frac{1}{\left(2t +x\right)^2}e^{-\frac{y^2}{4\left(2t+x\right)}},\\
\label{eq:Zhvcal1}
\frac{\partial\mathcal{M}_{(a)}}{\partial (p\cdot v)}|_{p=0} &=& \frac{g^2\tilde{\mu}^{2\epsilon}C_F}{(4\pi)^{d/2}} \int_0^{+\infty}dx\int_0^{+\infty}dy\, y\, \frac{1}{\left(2t +x\right)^{d/2}}e^{-\frac{y^2}{4\left(2t+x\right)}}.
\end{eqnarray}
Integrating out $x, y$, we obtain the results 
\begin{eqnarray}
\label{eq:deltamresult0}
-\mathcal{M}_{(a)}|_{p=0}&=& -i\frac{\alpha_s}{4\pi}C_F\frac{\sqrt{2\pi}}{\sqrt{t}},\\
\label{eq:Zhvresult0}
\frac{\partial\mathcal{M}_{(a)}}{\partial (p\cdot v)}|_{p=0} &=& - 2\times\frac{\alpha_sC_F}{4\pi}\left[\frac{1}{\epsilon_{\rm IR}}+\log\left(2\mu^2 te^{\gamma_E}\right)\right].
\end{eqnarray} 

Repeating the same calculation using dimensional regularization with $t=0$ gives vanishing $\mathcal{M}_{(a),\, t=0}|_{p=0}$ and 
\begin{eqnarray}
\frac{\partial\mathcal{M}_{(a),\, t=0}}{\partial (p\cdot v)}|_{p=0}=
2\frac{\alpha_sC_F}{4\pi}\left[\frac{1}{\epsilon_{\rm UV}}-\frac{1}{\epsilon_{\rm IR}}\right].
\end{eqnarray}
This confirms that we just need to subtract the IR poles from gradient flow results for the matching from the gradient-flow scheme to the $\overline{\rm MS}$ scheme when external momenta are set to zero.

Therefore, we obtain the following results (notice that the ``mass correction” in the renormalized Lagrangian is denoted as $i\delta m$)
\begin{eqnarray}
\label{eq:deltamresult}
\delta m &=&  -\frac{\alpha_s}{4\pi}C_R\frac{\sqrt{2\pi}}{\sqrt{t}}+ O(\alpha_s^2),\\
\label{eq:Zhvresult}
\zeta_{h_v}  &=& 1- \frac{\alpha_sC_R}{4\pi}\log\left(2\mu^2 te^{\gamma_E}\right) + O(\alpha_s^2),
\end{eqnarray}
where the $C_R =C_F$ for fundamental representation case and $C_R=C_A$ for adjoint representation case. The logarithmic dependence on $t$ in the above result of $\zeta_{h_v}$ is consistent with the renormalization constant of $h_v$ calculated in \cite{Craigie:1980qs}.

In the following matching calculations, we drop the calculations of un-flowed diagrams since they can be easily obtained from the amplitudes of the  corresponding flowed diagrams by setting $t=0$ from the beginning, which lead to results proportional to $\frac{1}{\epsilon_{\rm UV}} - \frac{1}{\epsilon_{\rm IR}}$ and obviously match the IR poles from flowed diagrams. We have explicitly checked that  the coefficients of $-\log(2\mu^2te^{\gamma_E})$ in the matching coefficients calculated below and above align with those of the UV poles  from the corresponding un-flowed operators at one-loop level , which can be served as a check point of our calculation.

\subsection{$\mathring{\zeta}_{\psi}$}
\begin{figure}
\centering
\includegraphics[width =0.72\textwidth]{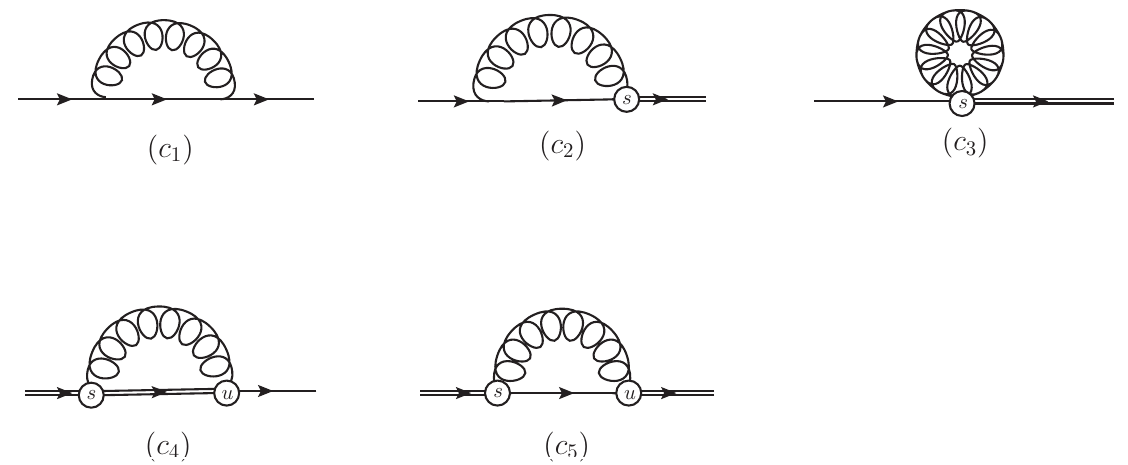}
\caption{One-loop quark self-energy Feynman diagrams in the gradient-flow formalism (mirror diagrams are implicit). The solid line is the fermion line, the double solid line is the flowed fermion line, the open circle represents  the flow vertex. }
\label{fig:quarkself}
\end{figure}
The Feynman diagrams for the one-loop corrections of the quark self-energy in the gradient-flow formalism are displayed in figure \ref{fig:quarkself}. A similar calculation was performed in ref.~\cite{Rizik:2020naq}, here we repeat the calculation for completeness, adhering to our strategy of splitting the matching coefficients into several parts that can be calculated separately. 

Diagram $(c_1)$ is the same as the one in ordinary QCD in the small flow-time limit, which gives the contribution 
\begin{eqnarray}
\label{eq:c1}
\frac{1}{i}\frac{\partial\mathcal{M}_{(c_1)}}{\partial \slashed{p}}|_{\slashed{p}=0}&=& -\frac{\alpha_sC_F}{4\pi}\left[\frac{1}{\epsilon_{\rm UV}}-\frac{1}{\epsilon_{\rm IR}}\right].
\end{eqnarray}
For diagrams $(c_2)$ and  $(c_3)$, we have 
\begin{eqnarray}
\label{eq:c2}
\mathcal{M}_{(c_2)}&=& 2\times (i)^2g^2\tilde{\mu}^{2\epsilon}C_F\int_0^tds\, \int\frac{d^dk}{(2\pi)^d}\frac{(-2\slashed{k})\slashed{k}}{\left(k^2 \right)^2}e^{-sk^2}\nonumber\\
&=&2\times\frac{\alpha_sC_F}{4\pi}\left[\frac{1}{\epsilon_{\rm UV}}+1+\log\left(2\mu^2 te^{\gamma_E}\right)\right],
\end{eqnarray}
and
\begin{eqnarray}
\label{eq:c3}
\mathcal{M}_{(c_3)}&=&2\times \frac{1}{2}g^2\tilde{\mu}^{2\epsilon}(-2dC_F)\int_0^tds\, \int\frac{d^dk}{(2\pi)^d}\frac{1}{k^2}e^{-2sk^2}\nonumber\\
&=&-4\times\frac{\alpha_sC_F}{4\pi}\left[\frac{1}{\epsilon_{\rm UV}}+\frac{1}{2}+\log\left(2\mu^2 te^{\gamma_E}\right)\right].
\end{eqnarray}

Diagram $(c_4)$ and its corresponding mirror diagram contribute proportionally to the external fermion momentum (with the gradient flow gauge parameter $\kappa=1$) and subsequently vanish upon setting the external momentum to zero. Diagram $(c_5)$ yields a null contribution due to its inherent oddness under the gluon loop momentum transformation $k\rightarrow -k$.

Adding up $\frac{1}{i}\frac{\partial\mathcal{M}_{(c_1)}}{\partial \slashed{p}}|_{\slashed{p}=0}$, $\mathcal{M}_{(c_2)}$ and $\mathcal{M}_{(c_3)}$, removing the UV divergences by introducing the ringed fermion field,  and subtracting the IR divergences through matching procedure, we obtain the one-loop matching coefficient for the matching from the ringed fermion field $\mathring{\chi}$ to the $\overline{\rm MS}$ renormalized fermion field $\psi$,
\begin{eqnarray}
\label{eq:quarkselfresult}
\mathring{\zeta}_{\psi}= 1 +  \frac{1}{2} \times\frac{\alpha_s}{4\pi}C_F\left[\log\left(2\mu^2 te^{\gamma_E}\right)-\log(432)\right] + O(\alpha_s^2).
\end{eqnarray}
\subsection{$\zeta_{\psi h_v}$}
\begin{figure}
\centering
\includegraphics[width =0.62\textwidth]{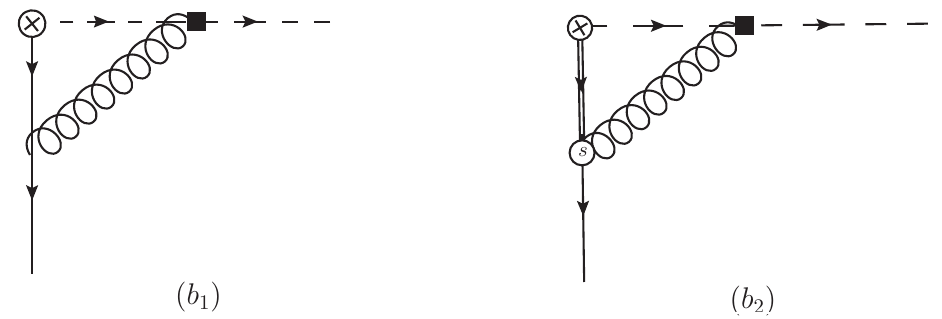}
\caption{Feynman diagrams of  the one-loop corrections for the $\bar{\psi}h_v$ vertex in gradient-flow formalism. The cross circle represents the $\bar{\psi}h_v$ vertex. The fermion field connected to the cross circle is flowed at  flow time $t$.}
\label{fig:qvertex}
\end{figure}
The one-loop correction Feynman diagrams for the $\bar{\psi}h_v$ vertex in gradient-flow formalism are given in figure \ref{fig:qvertex}. 
Setting the quark external momentum to be zero, we have the amplitude for diagram $(b_1)$,
\begin{eqnarray}
\label{eq:ZJstart}
\mathcal{M}_{(b_1)}= -g^2\tilde{\mu}^{2\epsilon}C_F\int\frac{d^dk}{(2\pi)^d}\frac{\left(\gamma\cdot v\right) \slashed{k}}{\left(-k\cdot v- i0\right)\left(k^2\right)^2} e^{-2tk^2}.
\end{eqnarray}
Obviously, the transverse components of $\slashed{k}$ give vanishing loop integration, thus we have
\begin{eqnarray}
\label{eq:ZJresult0}
\mathcal{M}_{(b_1)}&=&g^2\tilde{\mu}^{2\epsilon}C_F\int\frac{d^dk}{(2\pi)^d}\frac{1}{\left(k^2\right)^2} e^{-2tk^2}\nonumber\\
&=&\frac{\alpha_sC_F}{4\pi}\left[-\frac{1}{\epsilon_{\rm IR}}-\log\left(2\mu^2 te^{\gamma_E}\right) -1\right].
\end{eqnarray}

Diagram $(b_2)$ gives contribution proportional to the external  fermion momentum (with the  gradient flow gauge parameter $\kappa=1$) and subsequently vanishes upon setting the external momentum to zero.
Therefore, we have the one-loop result of  $\zeta_{\psi h_v}$ after matching to the $\overline{\rm MS}$ scheme,
\begin{eqnarray}
\label{eq:Zqhvresults}
\zeta_{\psi h_v}= 1 -\frac{\alpha_sC_F}{4\pi}\left[\log\left(2\mu^2 te^{\gamma_E}\right)+1\right] + O(\alpha_s^2).
\end{eqnarray} 
\subsection{$\zeta_A$}
\begin{figure}
\centering
\includegraphics[width =0.72\textwidth]{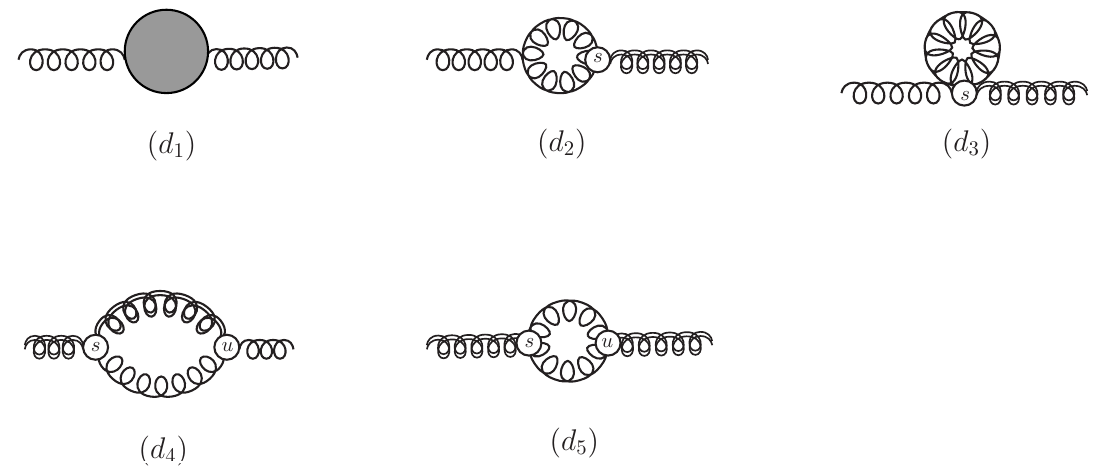}
\caption{Feynman diagrams for the one-loop gluon self-energy in gradient-flow formalism (mirror diagrams are implicit). The  blob in diagram $(d_1)$ represents the gluon vacuum  polarization. The single curvy lines are gluon lines, while the double curvy lines are the flowed gluon lines.}
\label{fig:gluonself}
\end{figure}
The Feynman diagrams for the one-loop corrections of gluon self-energy in the gradient-flow formalism are given by figure \ref{fig:gluonself}. For simplicity, we set the external momentum to be zero while extracting the corresponding contribution. Diagram $(d_1)$ is just the conventional gluon vacuum polarization, which gives the contribution in the small flow-time limit,
\begin{eqnarray}
\label{eq:d1}
\mathcal{M}_{(d_1)}&=&\frac{\alpha_s}{4\pi}\delta^{\mu\nu}\left(\frac{5}{3}C_A - \frac{2}{3}n_f\right)\left[\frac{1}{\epsilon_{\rm UV}}-\frac{1}{\epsilon_{\rm IR}}\right].
\end{eqnarray}

Diagrams $(d_2)$ and  $(d_3)$ give the contributions
\begin{eqnarray}
\label{eq:d2}
\mathcal{M}_{(d_2)}
&=& 4g^2C_A\tilde{\mu}^{2\epsilon}\int_0^t ds \int\frac{d^dk}{(2\pi)^d} \frac{(d-2)k^\mu k^\nu+k^2\delta^{\mu\nu}}{k^2 k^2}\text{exp}\left( -2s 
k^2 \right)\nonumber\\
&=&\frac{\alpha_s}{4\pi}C_A\delta^{\mu\nu}\left[\frac{3}{\epsilon_{\rm UV}} + 3\log\left(2\mu^2 te^{\gamma_E}\right)+ \frac{5}{2}\right],
\end{eqnarray}
and
\begin{eqnarray}
\label{eq:d3}
\mathcal{M}_{(d_3)}
&=& 2(1-d)g^2C_A\tilde{\mu}^{2\epsilon}\delta^{\mu\nu}\int_0^t ds \int\frac{d^dk}{(2\pi)^d} \frac{1}{k^2}\text{exp}\left( -2s 
k^2 \right)\nonumber\\
&=&-\frac{\alpha_s}{4\pi}C_A\delta^{\mu\nu}\left[\frac{3}{\epsilon_{\rm UV}} + 3\log\left(2\mu^2 te^{\gamma_E}\right)+ 1\right],
\end{eqnarray}
from which, we see that the UV divergences in diagrams $(d_2)$ and  $(d_3)$ cancel.

Diagram $(d_4)$ gives the contribution
\begin{eqnarray}
\label{eq:d4}
\mathcal{M}_{(d_4)} 
&=& 4g^2C_A\tilde{\mu}^{2\epsilon}\int_0^t ds\int_0^s du \int\frac{d^dk}{(2\pi)^d}\frac{2(d-2)k^\mu k^\nu+\delta^{\mu\nu}k^2}{k^2 }e^{-2sk^2}\nonumber\\
&=& \frac{\alpha_s}{4\pi}C_A\delta^{\mu\nu}\left[\frac{2}{\epsilon_{\rm UV}} + 2 \log\left(2\mu^2 te^{\gamma_E}\right)  -\frac{1}{2} \right].
\end{eqnarray}

Diagram $(d_5)$ does not contribute in the small flow-time limit because it gives contribution proportional to $t$ in the small flow-time expansion.
The results of $\mathcal{M}_{(d_2)}, \mathcal{M}_{(d_3)}$ and $\mathcal{M}_{(d_4)}$ are consistent with those given in ref.~\cite{Makino:2014taa}. 

Adding up $\mathcal{M}_{(d_1)}, \mathcal{M}_{(d_2)}, \mathcal{M}_{(d_3)}$ and $\mathcal{M}_{(d_4)}$, we obtain
\begin{eqnarray}
\label{eq:dsum}
\mathcal{M}_{(d)} 
&=&\mathcal{M}_{(d_1)} + \mathcal{M}_{(d_2)} + \mathcal{M}_{(d_3)} + \mathcal{M}_{(d_4)} \nonumber\\
&=& \frac{\alpha_s}{4\pi}\delta^{\mu\nu}\left[\frac{1}{\epsilon_{\rm UV}} \beta_0 - \frac{1}{\epsilon_{\rm IR}} \left(\frac{5}{3}C_A - \frac{2}{3}n_f\right)+ 2 C_A\log\left(2\mu^2 te^{\gamma_E}\right)  +C_A\right],
\end{eqnarray}
with $\beta_0=\frac{11}{3}C_A - \frac{2}{3}n_f$.
The IR poles are removed through matching procedure and the UV poles are absorbed into the renormalization of the strong coupling in the $\overline{\rm MS}$ scheme,
\begin{eqnarray}
Z_g = 1 - \frac{\alpha_s}{4\pi}\frac{1}{2\epsilon_{\rm UV}}\beta_0 + O(\alpha_s^2),
\end{eqnarray}
which is consistent with the general argument that the flowed gluon field in $B_{\mu}$ does not need renormalization while the strong coupling in $B_{\mu}$ needs to be renormalized \cite{Luscher:2011bx}.

Finally, we have the matching coefficient for the matching from the flowed $B_{\mu}$ field to the $\overline{\rm MS}$ renormalized gluon field $gA_{\mu}$, 
\begin{eqnarray}
\label{eq:ZAresults}
\zeta_A = 1+  \frac{\alpha_s}{4\pi}C_A\left[ \log\left(2\mu^2 te^{\gamma_E}\right)  +\frac{1}{2}\right] + O(\alpha_s^2).
\end{eqnarray}
Note that the logarithmic dependence in $\zeta_A $ cancel that in $\zeta_{h_v}$ (see eq.~(\ref{eq:Zhvresult})) in adjoint representation at one-loop level. However, the constant terms in the square brackets in eq.~(\ref{eq:ZAresults}) are not canceled.
\subsection{$\zeta_{\perp\perp}$ \& $\zeta_F$}
\begin{figure}
\centering
\includegraphics[width =0.8\textwidth]{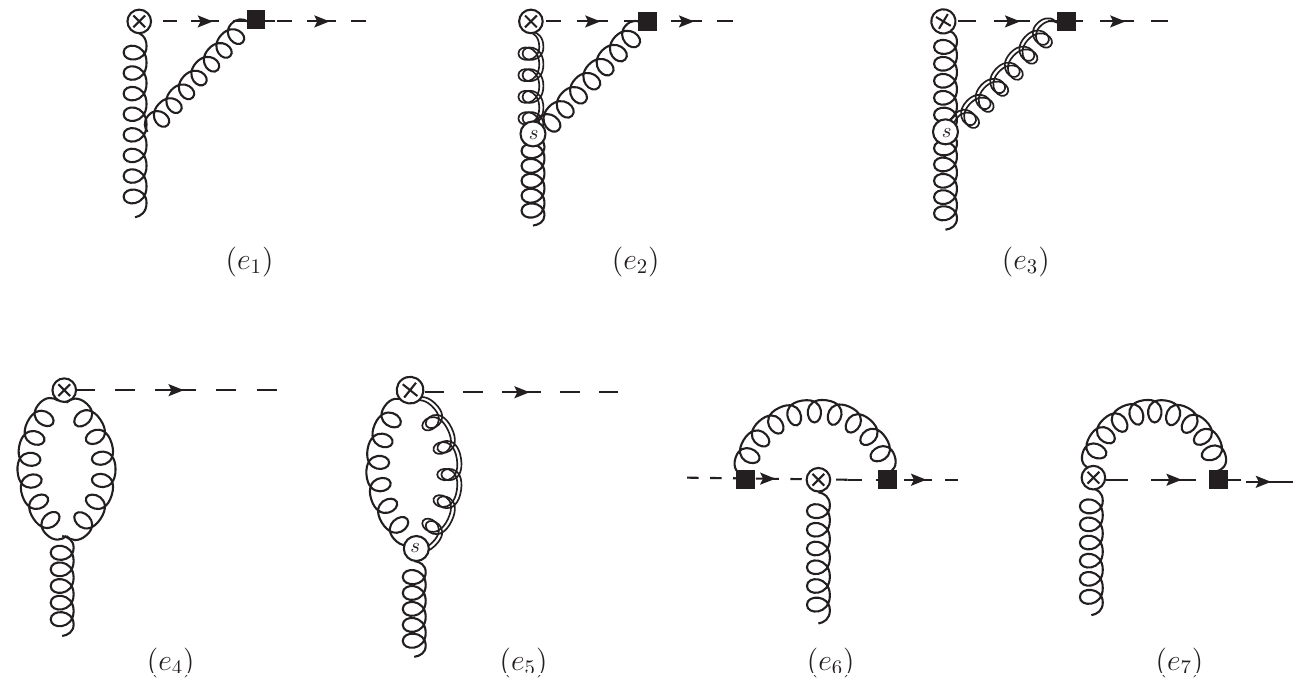}
\caption{Vertex type Feynman diagrams that contribute to the one-loop matching for the operators $gF^{\mu\nu}_{\|\perp}h_v,\, gF^{\mu\nu}_{\perp\perp}h_v$ and $g\bar{h}_vF^{\mu\nu}_{\perp\perp}h_v$ in the gradient-flow formalism (mirror diagrams are implicit). The gluon fields at the cross circles and the filled squares are flowed at flow time $t$. For $g\bar{h}_vF^{\mu\nu}_{\perp\perp}h_v$,  the fields $\bar{h}_v$ are not drawn in diagrams $(e_1) - (e_5)$  for simplicity. The  ``heavy” auxiliary fields in $g\bar{h}_vF^{\mu\nu}_{\perp\perp}h_v$ are in fundamental representation and the ``heavy” auxiliary fields in $gF^{\mu\nu}_{\|\perp}h_v,\, gF^{\mu\nu}_{\perp\perp}h_v$ are in adjoint representation. Diagrams $(e_6)$ and $(e_7)$ only give contributions to the operators $g\bar{h}_vF^{\mu\nu}_{\perp\perp}h_v$  and $gF^{\mu\nu}_{\|\perp}h_v$, respectively.}
\label{fig:gluonvertex}
\end{figure}
The one-loop vertex type Feynman diagrams for the operators $gF^{\mu\nu}_{\|\perp}h_v,\, gF^{\mu\nu}_{\perp\perp}h_v$ and $g\bar{h}_vF^{\mu\nu}_{\perp\perp}h_v$ in the gradient-flow formalism are shown in figure \ref{fig:gluonvertex}. Choosing the external gluon momentum to be $q$ (flowing into the vertex), we have the leading-order amplitudes 
\begin{eqnarray}
\mathcal{M}_{F_{\|\perp}h_v}^{\rm LO} &=& ig\left(q_{\|}^{\mu}A^{\nu}_{\perp} -q_{_\perp}^{\nu}A^{\mu}_{\|} \right),\\
\mathcal{M}_{F_{\perp\perp }h_v}^{\rm LO} &=& ig\left(q_{\perp}^{\mu}A^{\nu}_{\perp} -q_{\perp}^{\nu}A^{\mu}_{\perp} \right),\\
\mathcal{M}_{\bar{h}_vF_{\perp\perp }h_v}^{\rm LO} &=& ig\left(q_{\perp}^{\mu}A^{\nu}_{\perp} -q_{\perp}^{\nu}A^{\mu}_{\perp} \right),
\end{eqnarray}
where we have suppressed the color indices and dropped the external ``heavy” fields but kept the external gluon fields for the convenience of matching procedure.

Since we are only interested in obtaining the matching coefficients, which are independent of the external momenta, we can set $q\cdot v=0$ for the amplitudes of the operators $gF^{\mu\nu}_{\perp\perp}h_v$ and $g\bar{h}_vF^{\mu\nu}_{\perp\perp}h_v$, but not for the amplitudes of the operator $gF^{\mu\nu}_{\|\perp}h_v$.
The computations of the one-loop vertex corrections for the operators  $gF^{\mu\nu}_{\perp\perp}h_v$ and $g\bar{h}_vF^{\mu\nu}_{\perp\perp}h_v$ are straightforward and similar. These three operators share the same diagrams $(e_1) - (e_5)$. We will only give the amplitudes of  diagrams $(e_1) - (e_5)$ for the operators $gF^{\mu\nu}_{\|\perp}h_v$ and $gF^{\mu\nu}_{\perp\perp}h_v$, supplied with the amplitude of diagram $(e_6)$ for the operator $g\bar{h}_vF^{\mu\nu}_{\perp\perp}h_v$ and the amplitude of diagram $(e_7)$ for the operator $g\bar{h}_vF^{\mu\nu}_{\|\perp}h_v$.  

The calculation of the vertex corrections for $gF^{\mu\nu}_{\|\perp}h_v$ is not so straightforward due to non-trivial mixing with the gauge non-invariant operator $iv^{\mu}A_{\perp}^{\nu}(iv\cdot D - i\delta m) h_v$.
We will deal with $gF^{\mu\nu}_{\|\perp}h_v$ in the next subsection in details. 

The amplitudes of diagrams $(e_1) - (e_6)$ are given by
\begin{eqnarray}
\label{eq:e1}
\mathcal{M}_{(e_1)} 
&=&-g^3C_A\tilde{\mu}^{2\epsilon}\int\frac{d^dk}{(2\pi)^d}\bigg\{ik^{\mu}A^{\alpha}\frac{\delta^{\nu\alpha}\left(-q-k\right)\cdot v+v^{\alpha}\left(2q-k\right)^{\nu}+v^{\nu}\left(2k-q\right)^{\alpha}}{(k\cdot v- i0)k^2(k-q)^2}\nonumber\\
&&\, \, \, \, \, \, \, \, \, \, \, \, \, \, \, \, \, \, \, \, \, \, \, \, \, \, \, \, \, \, \, \, \, \, \, \, \, \, \, \, \, \, \, \, \, \, \, \, \, \, \, \, \,  \, \, \, \, \, \, -\left(\mu\leftrightarrow\nu\right)\bigg\}e^{-t\left[k^2+(k-q)^2\right]},
\end{eqnarray}
\begin{eqnarray}
\label{eq:e2}
\mathcal{M}_{(e_2)} 
&=&g^3C_A\tilde{\mu}^{2\epsilon}\int\frac{d^dk}{(2\pi)^d}\int_0^t ds\bigg\{ik^{\mu}\frac{v^{\alpha}\left(k-2q\right)^{\nu}+2\delta^{\alpha \nu}q\cdot v-2v^{\nu}(k-q)^{\alpha}}{(k\cdot v- i0)(k-q)^2}\nonumber\\
&&\, \, \, \, \, \, \, \, \, \, \, \, \, \, \, \, \, \, \, \, \, \, \, \, \, \, \, \, \, \, \, \, \, \, \, \, \, \, \, \, \, \, \, \, \, \, \, \, \, \, \, \, \,  \, \, \, \, \, \, \, \, \, \, \, \, \, \,  \, \, \, -\left(\mu\leftrightarrow\nu\right)\bigg\}A^{\alpha}e^{-(s+t)(k-q)^2-(t-s)k^2},\\\nonumber\\
\label{eq:e3}
\mathcal{M}_{(e_3)} 
&=&g^3C_A\tilde{\mu}^{2\epsilon}\int\frac{d^dk}{(2\pi)^d}\int_0^t ds\bigg\{ik^{\mu}\frac{\delta^{\nu\alpha}\left(q+k\right)\cdot v+2v^{\nu}(-k)^{\alpha}-2v^{\alpha}q^{\nu}}{(k\cdot v- i0)k^2}\nonumber\\
&&\, \, \, \, \, \, \, \, \, \, \, \, \, \, \, \, \, \, \, \, \, \, \, \, \, \, \, \, \, \, \, \, \, \, \, \, \, \, \, \, \, \, \, \, \, \, \, \, \, \, \, \, \, \, \, \, \, \, \, \, \, \, \, \, \, \, \, \, \, \,  -\left(\mu\leftrightarrow\nu\right)\bigg\}A^{\alpha}e^{-(t+s)k^2-(t-s)(k-q)^2},\\
\label{eq:e4}
\mathcal{M}_{(e_4)} 
&=&-\frac{3}{2}g^3C_A\tilde{\mu}^{2\epsilon}\int\frac{d^dk}{(2\pi)^d } \frac{i\left(q^{\mu}A^{\nu}-q^{\nu}A^{\mu}\right)}{k^2(k-q)^2}e^{-t\left[k^2+(k-q)^2\right]},
\end{eqnarray}
\begin{eqnarray}
\label{eq:e5}
\mathcal{M}_{(e_5)} 
&=&-g^3C_A\tilde{\mu}^{2\epsilon}\int\frac{d^dk}{(2\pi)^d }\int_0^t ds\frac{i(k+3q)^{\mu}A^{\nu}-i(k+3q)^{\nu}A^{\mu}}{k^2}e^{-(t+s)k^2-(t-s)(k-q)^2},\nonumber\\\\
\mathcal{M}_{(e_6)} 
&=&g^3\left(C_F-\frac{C_A}{2}\right)\tilde{\mu}^{2\epsilon}\int\frac{d^dk}{(2\pi)^d }\int_0^t ds\frac{i\left(q^{\mu}_{\perp}A^{\nu}_{\perp}-q^{\nu}_{\perp}A^{\mu}_{\perp}\right)}{(k\cdot v-i0)^2k^2}e^{-2tk^2}.
\end{eqnarray}
Extracting the LO expressions and setting $q=0$ thereafter, we obtain the contributions from each diagram for the operator $gF^{\mu\nu}_{\perp\perp}h_v$,
\begin{eqnarray}
\label{eq:e1perpperpsimresult}
\zeta_{\perp\perp,\, (e_1)}&=&
-\frac{\alpha_sC_A}{4\pi}\times \frac{1}{2} \left[\frac{1}{\epsilon_{\rm IR}}+\log\left(2\mu^2 te^{\gamma_E}\right) +1\right],\\
\label{eq:e2perpperp}
\zeta_{\perp\perp,\, (e_2)}
&=&0,\\
\label{eq:e3perpperpresult}
\zeta_{\perp\perp,\, (e_3)}
&=&\frac{\alpha_s}{4\pi}C_A\times\frac{1}{16},\\
\label{eq:e4perpperpresult}
\zeta_{\perp\perp,\, (e_4)}&=&\frac{\alpha_sC_A}{4\pi}\times \frac{3}{2} \left[\frac{1}{\epsilon_{\rm IR}}+\log\left(2\mu^2 te^{\gamma_E}\right) +1\right],\\
\label{eq:e5perperpresult}
\zeta_{\perp\perp,\, (e_5)}
&=&- \frac{\alpha_sC_A}{4\pi}\times\frac{25}{16}.
\end{eqnarray}
Summing up all the contributions from diagrams $(e_1) - (e_5)$ and discarding the IR poles through matching procedure, we obtain the vertex matching coefficient for the operator $gF^{\mu\nu}_{\perp\perp}h_v$,
\begin{eqnarray}
\label{eq:resultgluonvertexperpperp}
\zeta_{\perp\perp} &=& 1+  \frac{\alpha_sC_A}{4\pi}\left[\log\left(2\mu^2 te^{\gamma_E}\right) -\frac{1}{2}\right]+ O(\alpha_s^2).
\end{eqnarray}
Similarly, we have the contributions from each diagram for the operator $g\bar{h}_vF^{\mu\nu}_{\perp\perp}h_v$,
\begin{eqnarray}
\zeta_{F,\, (e_1)}   &=& - \frac{\alpha_s}{4\pi}C_A\times \frac{1}{2}\left[\frac{1}{\epsilon_{\rm IR}} +\log\left(2\mu^2 te^{\gamma_E}\right) + 1\right],\\
\zeta_{F,\, (e_2)}   &=& 0,\\
\zeta_{F,\, (e_3)} &=&\frac{\alpha_s}{4\pi}C_A\times \frac{1}{16},\\
\zeta_{F,\, (e_4)}   &=&  \frac{\alpha_s}{4\pi}C_A\times \frac{3}{2}\left[\frac{1}{\epsilon_{\rm IR}} +\log\left(2\mu^2 te^{\gamma_E}\right) + 1\right],\\
\zeta_{F,\, (e_5)}   &=& - \frac{\alpha_s}{4\pi}C_A\times \frac{25}{16},\\
\zeta_{F,\, (e_6)}  &= & \frac{\alpha_s}{4\pi}\left(C_F -\frac{C_A }{2}\right)\times  2\left[\frac{1}{\epsilon_{\rm IR}} +\log\left(2\mu^2 te^{\gamma_E}\right)\right],
\end{eqnarray}  
which lead to 
\begin{eqnarray}
\label{eq:resultgluonvertexF}
\zeta_{\perp\perp}^F&= &1+ \frac{\alpha_s}{4\pi} \left[2C_F\log\left(2\mu^2 te^{\gamma_E}\right)+\frac{1}{2}C_A\right] + O(\alpha_s^2).
\end{eqnarray}
\subsection{$\zeta_{\|\perp}$}
Diagrams $(e_4), (e_5)$ have no mixing and give the same contributions for $gF^{\mu\nu}_{\perp\perp}h_v$ and $gF^{\mu\nu}_{\|\perp}h_v$, 
\begin{eqnarray}
\label{eq:e4pperpresult}
\zeta_{\|\perp,\, (e_4)}&=&\frac{\alpha_sC_A}{4\pi}\times \frac{3}{2} \left[\frac{1}{\epsilon_{\rm IR}}+\log\left(2\mu^2 te^{\gamma_E}\right) +1\right],\\
\label{eq:e5pperpresult}
\zeta_{\|\perp,\, (e_5)}
&=&- \frac{\alpha_sC_A}{4\pi}\times\frac{25}{16}.
\end{eqnarray}
\\
Complications arise from diagrams $(e_1), (e_2), (e_3)$ and $(e_7)$, whose amplitudes are given by
\begin{eqnarray}
\label{eq:e1pperp}
\mathcal{M}_{(e_1)}^{\|\perp}
&=&ig^3C_A\tilde{\mu}^{2\epsilon}\int\frac{d^dk}{(2\pi)^d}\left[\frac{(q+k)^{\mu}_{\|} A_{\perp}^{\nu}-(2q-k)^{\nu}_{\perp} A^{\mu}_{\|}}{k^2(k-q)^2} +\frac{v^{\mu}k_{\perp}^{\nu}(2k-q)_{\perp}\cdot A_{\perp} }{(k\cdot v- i0)k^2(k-q)^2} \right]\nonumber\\
&&\, \, \, \, \, \, \, \, \, \, \, \, \, \, \, \, \, \, \, \, \, \, \, \, \, \, \, \, \, \, \, \, \, \, \, \, \, \, \, \, \, \, \, \, \, \, \, \, \, \,  \, \, \, \, \, \, \, \, \, \, \, \,  \, \, \, \, \, \, \, \, \, \, \, \,  \, \, \times e^{-t\left[k^2+(k-q)^2\right]},
\end{eqnarray}
\begin{eqnarray}
\label{eq:e2pperp}
\mathcal{M}_{(e_2)}^{\|\perp} 
&=&ig_s^3C_A\tilde{\mu}^{2\epsilon}\int\frac{d^dk}{(2\pi)^d}\int_0^t ds \left[\frac{2\left(q^{\mu}_{\|}A^{\nu}_{\perp} - q_{\perp}^{\nu}A_{\|}^{\mu}\right)}{(k-q)^2} +\frac{2v^{\mu}k_{\perp}^{\nu}(k-q)\cdot A}{(k\cdot v-i0)(k-q)^2}\right]\nonumber\\
&&\, \, \, \, \, \, \, \, \, \, \, \, \, \, \, \, \, \, \, \, \, \, \, \, \, \, \, \, \, \, \, \, \, \, \, \, \, \, \, \, \, \, \, \, \, \, \, \, \, \,  \, \, \, \, \, \, \, \, \, \, \, \,  \, \, \, \, \, \, \, \, \, \, \, \,  \, \, \times e^{-(s+t)(k-q)^2-(t-s)k^2},
\end{eqnarray}
\begin{eqnarray}
\label{eq:e3pperp}
\mathcal{M}_{(e_3)}^{\|\perp}
&=&ig_s^3C_A\tilde{\mu}^{2\epsilon}\int\frac{d^dk}{(2\pi)^d}\int_0^t ds \Bigg[\frac{(k+q)^{\mu}_{\|}A^{\nu}_{\perp}-(k+q)^{\nu}_{\perp}A^{\mu}_{\|}}{k^2}\nonumber\\
&&+\frac{(A\cdot v)\left(q^{\mu}_{\|}k^{\nu}_{\perp}-q^{\nu}_{\perp}k^{\mu}_{\|}\right)+2v^{\mu}k_{\perp}^{\nu}(k\cdot A)}{(k\cdot v-i0)k^2}\Bigg]e^{-(s+t)k^2-(t-s)(k-q)^2},
\end{eqnarray}
\begin{eqnarray}
\label{eq:e7pperp}
\mathcal{M}_{(e_7)}^{\|\perp}
&=&ig^3C_A\tilde{\mu}^{2\epsilon}\int\frac{d^dk}{(2\pi)^d}\frac{-v^{\mu}A_{\perp}^{\nu} }{(k\cdot v-i0)(k-q)^2}e^{-2t(k-q)^2}.
\end{eqnarray}
These amplitudes do not give expressions that are exactly proportional to the LO expression of the operator, $gF^{\mu\nu}_{\|\perp}h_v$ even after tensor reduction for the loop integrations. They lead to mixing with the gauge non-invariant operators $iv^{\mu}A_{\perp}^{\nu}(iv\cdot D - i\delta m) h_v$, whose LO expression is proportional to $q_{\|}^{\mu}A_{\perp}^{\nu}$, and thus it is not straightforward to disentangle  the mixing from those that give contributions to the operator $gF^{\mu\nu}_{\|\perp}h_v$. Fortunately, we can extract the contributions to the operator $gF^{\mu\nu}_{\|\perp}h_v$ through the results that are proportional to $q_{\perp}^{\nu}A_{\|}^{\mu}$. To check the mixing pattern and cancellation of linear divergences that were found in refs.~\cite{Dorn:1981wa,Zhang:2018diq,Wang:2019tgg}, we will also keep the results that are proportional to $q_{\|}^{\mu}A_{\perp}^{\nu}$ and $v_{\|}^{\mu}A_{\perp}^{\nu}/\sqrt{t}$ in the small flow-time limit. 

If we set $t=0$ in $\mathcal{M}_{(e_1)}^{\|\perp}$ and $\mathcal{M}_{(e_7)}^{\|\perp}$, we can easily see that the linear divergences coming from $\mathcal{M}_{(e_7)}^{\|\perp}$ and the second term of $\mathcal{M}_{(e_1)}^{\|\perp}$ cancel with $d=3-2\epsilon$. However, in gradient  flow scheme, the cancellation of linear divergences are much more complicated. Apparently, the  linearly divergences arising from $\mathcal{M}_{(e_7)}^{\|\perp}$ and the second term of $\mathcal{M}_{(e_1)}^{\|\perp}$ do not cancel with $d=4-2\epsilon$ and $t>0$. The cancellation of linear divergences in gradient-flow scheme will be restored when diagrams $(e_2)$ and $(e_3)$ are included.

Great simplification can be achieved for our matching calculation by adopting the well-developed method used in the matching calculation in HQET. That is, we focus on the large loop momentum region $k >> q$ and expand the amplitudes up to linear order of the external momentum $q$ before performing the loop momentum integration.
This method is valid in our case because, in the small flow-time limit, the flowed operators share the same IR behavior with the corresponding un-flowed operators and contributions to the matching coefficients come from the UV region.

Expanding the amplitudes according to $k >> q$ up to linear order of $q$ and completing the loop integrations, we obtain
\begin{eqnarray}
\label{eq:e1pperpexpand}
\mathcal{M}_{(e_1)}^{\|\perp}
&=& \frac{\alpha_s}{4\pi}gC_A\left\{ -\frac{2}{3}\frac{\sqrt{2\pi}}{\sqrt{t}}v^{\mu}A_{\perp}^{\nu} - i \ell_1 \left[\left(\frac{3}{2} + \frac{4}{d}\right)q^{\mu}_{\|} A_{\perp}^{\nu}-\frac{3}{2}q^{\nu}_{\perp} A^{\mu}_{\|}\right]\right\},
\end{eqnarray}
\begin{eqnarray}
\label{eq:e2pperpexpand}
\mathcal{M}_{(e_2)}^{\|\perp} 
&=& \frac{\alpha_s}{4\pi}gC_A\left\{ -\frac{1}{6}\frac{\sqrt{2\pi}}{\sqrt{t}}v^{\mu}A_{\perp}^{\nu} + i \left[\frac{15}{8}q^{\mu}_{\|} A_{\perp}^{\nu}-\frac{1}{8}q^{\nu}_{\perp} A^{\mu}_{\|}\right]\right\},
\end{eqnarray}
\begin{eqnarray}
\label{eq:e3pperpexpand}
\mathcal{M}_{(e_3)}^{\|\perp}
&=& \frac{\alpha_s}{4\pi}gC_A\left\{ -\frac{1}{6}\frac{\sqrt{2\pi}}{\sqrt{t}}v^{\mu}A_{\perp}^{\nu} + i \left[\frac{11}{16}q^{\mu}_{\|} A_{\perp}^{\nu}-\frac{15}{16}q^{\nu}_{\perp} A^{\mu}_{\|}\right]\right\},
\end{eqnarray}
\begin{eqnarray}
\label{eq:e7pperpexpand}
\mathcal{M}_{(e_7)}^{\|\perp}
&=& \frac{\alpha_s}{4\pi}gC_A\left\{\frac{\sqrt{2\pi}}{\sqrt{t}}v^{\mu}A_{\perp}^{\nu} + \frac{i}{2} \ell_1\frac{1}{2}q^{\mu}_{\|} A_{\perp}^{\nu}\right\},
\end{eqnarray}
where 
\begin{eqnarray}
\ell_1 = \frac{1}{\epsilon_{\rm IR}}+\log\left(2\mu^2 te^{\gamma_E}\right) +1.
\end{eqnarray}
From the above results, we can clearly see the cancellation of the linear divergences. After subtracting the contributions to the operator $gF^{\mu\nu}_{\|\perp}h_v$, the leftover logarithmic dependence on $t$ also matches the UV divergence of the mixing presented in refs.~\cite{Dorn:1981wa,Zhang:2018diq,Wang:2019tgg}.

Summarizing all the contributions from diagrams $(e_1) - (e_5)$ and $(e_7)$, we obtain the matching coefficient contributed from the vertex correction of the operator $gF^{\mu\nu}_{\|\perp}h_v$,
\begin{eqnarray} 
\label{eq:resultgluonvertexpperp}
\zeta_{\|\perp} &=& 1 -\frac{1}{2}\frac{\alpha_s}{4\pi}C_A+ O(\alpha_s^2),
\end{eqnarray}
which cancels the constant term in the square brackets of $\zeta_{A}$ in eq.~(\ref{eq:ZAresults}) and thus makes the one-loop contribution to the matching coefficient of the operator $gF^{\mu\nu}_{\|\perp}h_v$ vanish \footnote{In ref.~\cite{Eller:2021qpp}, the author calculated the color-electric field correlator $G_E$ at one-loop level within the gradient-flow formalism. The matching calculation for $G_E$ in the small flow-time limit can be reduce to the matching calculation for the local current operator $g\bar{h}_vF^{\mu\nu}_{\|\perp}h_v$ with $h_v$ being in fundamental representation, which also has vanishing one-loop contribution to the matching coefficient.}.

\subsection{Results and applications for Wilson-line operators}
We are now ready to summarize the final results of the matching coefficients for the four local current operators: $\bar{\psi} h_v$, $gF^{\mu\nu}_{\|\perp}h_v$, $gF^{\mu\nu}_{\perp\perp}h_v$ and $g\bar{h}_vF^{\mu\nu}_{\perp\perp}h_v$. The relationships between the matching coefficients of the local current operators and the matching coefficients of the fields and vertices are provided in eqs.~(\ref{eq:operatorrelations}).
The one-loop results of $\zeta_{h_v}$, $\mathring{\zeta}_{\psi}$,  $\zeta_{\psi h_v}$, $\zeta_{A}$, $\zeta_{\|\perp}$,  $\zeta_{\perp\perp}$ and $\zeta_F$ can be found in eq.~(\ref{eq:Zhvresult}), eq.~(\ref{eq:quarkselfresult}), eq.~(\ref{eq:Zqhvresults}), eq.~(\ref{eq:ZAresults}), eq.~(\ref{eq:resultgluonvertexpperp}), eq.~(\ref{eq:resultgluonvertexperpperp}) and eq.~(\ref{eq:resultgluonvertexF}), respectively. Implementing those results, we finally obtain
\begin{eqnarray}
\label{eq:cpsihvfinal}
c_{\psi h_v}(t, \mu) &=& 1- \frac{\alpha_s}{4\pi}C_F\left[\frac{3}{2}\log\left(2\mu^2 te^{\gamma_E}\right)+\frac{\log(432)}{2}+1\right] +O(\alpha_s^2),\\
\label{eq:cpperpfinal}
c_{\|\perp} (t, \mu) &=& 1+O(\alpha_s^2),\\
\label{eq:cperpperpfinal}
c_{\perp\perp} (t, \mu) &=&1+ \frac{\alpha_s}{4\pi}C_A\times\log\left(2\mu^2 te^{\gamma_E}\right) +O(\alpha_s^2),\\
\label{eq:cpperpFfinal}
c_F(t, \mu)  &=& 1+\frac{\alpha_s}{4\pi}C_A\times\log\left(2\mu^2 te^{\gamma_E}\right) +O(\alpha_s^2).
\end{eqnarray}
We have  verified that the logarithmic dependences on $t$ in eq.~(\ref{eq:cpsihvfinal}), eq.~(\ref{eq:cpperpfinal}) and eq.~(\ref{eq:cperpperpfinal}) align with the anomalous dimensions given by refs.~\cite{Dorn:1980hs,Dorn:1981wa,Dorn:1986dt} for the current operators $\bar{\psi} h_v$, $gF^{\mu\nu}_{\|\perp}h_v$ and $gF^{\mu\nu}_{\perp\perp}h_v$, respectively (see ref.~\cite{Braun:2020ymy}  that uses similar notation for these operators for comparison). For the current operator $g\bar{h}_vF^{\mu\nu}_{\perp\perp}h_v$, its anomalous dimension  was also calculated in refs.~\cite{Eichten:1990vp,Falk:1990pz} using background field method (see ref.~\cite{Abbott:1981ke}  for a review of background field method), which is consistent with the logarithmic dependences on $t$  in eq.~(\ref{eq:cpperpFfinal}). It is noteworthy that while $gA$ does not contribute to the anomalous dimension of $g\bar{h}_vF^{\mu\nu}_{\perp\perp}h_v$ within the background field method, it does so using the method adopted in this work. The UV poles in the $\overline{\rm MS}$ scheme for diagrams $(e_1)$ and $(e_4)$ differ depending on whether the calculations are conducted with or without the background field method, owing to the distinct Feynman rules of three-gluon vertex. We have checked that these differences compensate for those arising from the anomalous dimensions of $gA$ with or without the background field method using Feynman gauge. With  correct logarithmic dependences on $t$ for the matching coefficients of  local current operators,  ``heavy” auxiliary field $h_v$,  quark field $\psi$ and $gA$, the corresponding logarithmic dependences on $t$ for the vertex matching coefficients are ensured to be correct. While the results of $c_{\perp\perp}(t, \mu)$ and $c_F(t, \mu)$ coincide at  one-loop level, it is intriguing that  the anomalous dimensions of $gF^{\mu\nu}_{\perp\perp}h_v$ and $g\bar{h}_vF^{\mu\nu}_{\perp\perp}h_v$ differ by a term proportional to $\pi^2$ at  two-loop level (see the two-loop anomalous dimensions given in refs.~\cite{Braun:2020ymy, Amoros:1997rx,Czarnecki:1997dz}). In addition, for the operator $gF^{\mu\nu}_{\|\perp}h_v$, its anomalous dimension starts to contribute at two-loop level \cite{Braun:2020ymy}, the two-loop and three-loop anomalous dimensions of the operator $\bar{\psi} h_v$ can be found in refs.~\cite{Broadhurst:1991fz,Ji:1991pr} and ref.~\cite{Chetyrkin:2003vi}, respectively.

As mentioned in the end of section \ref{subsec:smallflowexp}, the matching coefficients for the Wilson-line operators can be directly obtained from the corresponding matching coefficients of the local current operators in the small flow-time limit within the one-dimensional auxiliary-field formalism. In the small flow-time limit, expressing the matching equations for the Wilson-line operators defined in eq.~(\ref{eq:defqoperator}), eq.~(\ref{eq:defpperp}) and eq.~(\ref{eq:defperpperp}) as
\begin{eqnarray}
\mathcal{O}_{\Gamma}^{\rm R}(zv, t) &=& \mathcal{C}_{\psi}(t, \mu)e^{\delta m z}\mathcal{O}_{\Gamma}^{\overline{\rm MS}}(zv) + O(t) ,\\
\mathcal{O}_{\|\perp}^{\rm R} (zv, t) &=&  \mathcal{C}_{\|\perp}(t, \mu)e^{\delta m z}\mathcal{O}_{\|\perp}^{\overline{\rm MS}} (zv)+ O(t),\\ 
\mathcal{O}_{\|\perp}^{\rm R} (zv, t) &=& \mathcal{C}_{\perp\perp}(t, \mu)e^{\delta m z}\mathcal{O}_{\perp\perp}^{\overline{\rm MS}} (zv)+ O(t),
\end{eqnarray}
where we have omitted the Lorentz indices for simplicity, we establish relations
\begin{eqnarray}
\mathcal{C}_{\psi}(t, \mu) &=& c^2_{\psi h_v}(t, \mu),\\
\mathcal{C}_{\|\perp}(t, \mu) &=& c^2_{\|\perp} (t, \mu),\\
\mathcal{C}_{\perp\perp}(t, \mu) &=& c^2_{\perp\perp} (t, \mu).
\end{eqnarray}
These relations are independent of $z$ and external states of the matrix elements constructed using these Wilson-line operators. 

Regarding the spin-dependent potentials in terms of chromomagnetic field insertions into a Wilson loop, each such insertion will lead to a factor of the matching coefficient $c_F(t, \mu)$ for the matching from the gradient-flow scheme to the $\overline{\rm MS}$ scheme.

\section{Summary and outlook}
\label{sec:summary}
We have developed a systematic approach for matching from the gradient-flow scheme to the $\overline{\rm MS}$ scheme for Wilson-line operators in the small flow-time limit. This approach extends to various applications, including quasi-PDFs, gluonic correlators that appear in quarkonium decay and production in the framework of pNRQCD and spin-dependent potentials in terms of chromomagnetic field insertions into a Wilson loop.

Within the one-dimensional auxiliary-field formalism, the matching from the gradient-flow scheme to the $\overline{\rm MS}$ scheme for the Wilson-line operators discussed in this work can be reduced to the matching for the corresponding local current operators in the small flow-time limit.
We match these local current operators from the gradient-flow scheme to the $\overline{\rm MS}$ scheme by comparing perturbative computations in these two schemes.
Great simplification for the matching procedure can be achieved by using small flow-time expansion and setting external momenta to be zero whenever possible.
In this way, the well-developed method of matching calculation in HQET can be applied. After completing the matching for the local current operators, one automatically obtains the matching coefficients for the corresponding Wilson-line operators in the small flow-time limit,
because the combinations of the renormalized local current operators at different spacetime do not introduce further UV divergences, apart from the subtracted linear divergences, due to the multiplicative renormalizability of the Wilson-line operators.

The matching for the local current operators can be further decomposed as the matching for the fields and vertices due to the multiplicative renormalizability of the local current operators.
We thus compute the matching coefficients at one-loop level for auxiliary ``heavy” auxiliary field $h_v$,  massless quark field $\psi$, $\psi h_v$ vertex,  gluon field $gA$ and the vertices of the local current operators  $gF^{\mu\nu}_{\|\perp}h_v,\, gF^{\mu\nu}_{\perp\perp}h_v$ and $g\bar{h}_vF^{\mu\nu}_{\perp\perp}h_v$. The one-loop result of the linearly divergent ``mass” correction $\delta m$ is also obtained. Provided with these results, we give the matching coefficients for the local current operators and the Wilson-line operators in the small flow-time limit. 

It is worth mentioning that this method is also applicable for matching from lattice regularization schemes to the $\overline{\rm MS}$ scheme in the small lattice spacing limit, albeit with added complexity due to UV and IR divergence treatment \cite{Constantinou:2017sej}.

In the appendix, we have shown that, at one-loop level, for the case of hadronic matrix element defined in eq.~(\ref{eq:quarkhmte}) at zero external momenta, the finite flow time effect is very small as long as the flow radius is smaller than physical distance $z$, which is usually satisfied in lattice gradient flow computations.

With the systematic approach we have developed for the matching from the gradient-flow scheme to the $\overline{\rm MS}$ scheme for the Wilson-line operators, 
 it is straightforward to extend the current study to two-loop order, which would be more helpful for lattice gradient flow computations of matrix elements constructed using the off-lightcone Wilson-line operators. We leave it for future research.
 
\subsubsection*{Note added}
Before submission of this work, we noticed that the one-loop result of the matching coefficient for the gluonic correlator $G_B$ was recently published in ref.~\cite{Altenkort:2023eav},  which was the same as our one-loop result of  $c_F^2(t, \mu)$, because the matching calculation for the gluonic correlator $G_B$ could be reduced to the matching calculation for the local current operator $g\bar{h}_vF^{\mu\nu}_{\perp\perp}h_v$.

\acknowledgments 
We would like to thank X. Ji,  Y. Ji, C. Monahan, T.  Neumann, A. Shindler, V. Harlander, F. Lange and P. Petreczky for useful discussions. We thank A. Vairo for reading the paper and giving very useful comments. The work of N.~B. and X.-P.~W. is supported by the DFG (Deutsche Forschungsgemeinschaft,
German Research Foundation) Grant No. BR 4058/2-2. We
acknowledge support from the DFG cluster of excellence ``ORIGINS'' under
Germany's Excellence Strategy - EXC-2094 - 390783311. The authors acknowledge support from STRONG-2020- European Union’s Horizon 2020 research and innovation program under grant agreement
No. 824093.

\appendix
\section{Quark quasi-PDF operator  in gradient flow}
\label{app:quarkquasi}
\begin{figure}
\centering
\includegraphics[width =0.72\textwidth]{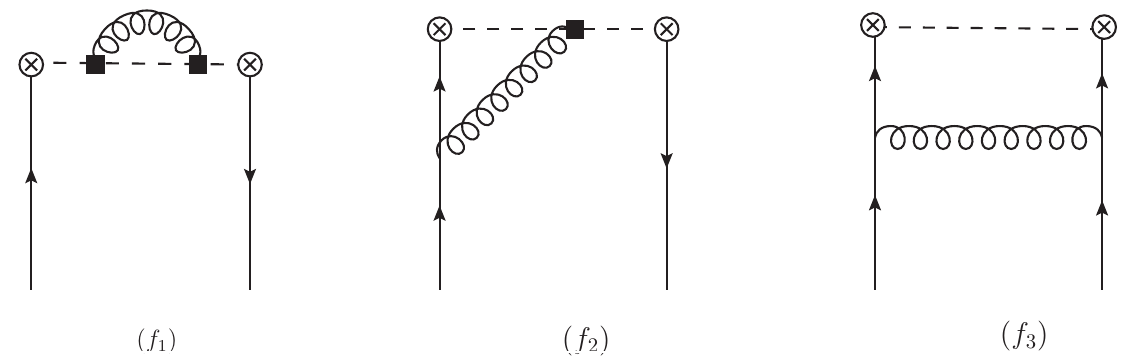}
\caption{Feynman diagrams  that contribute to the one-loop corrections for the hadronic matrix element defined as eq.~(\ref{eq:quarkhmte}) in gradient-flow formalism. The dashed line is the Wilson line, the filled squares represent the flowed $B_{\mu}$ field at flow time $t$, the cross circles represent the spacetime points that connecting the Wilson line and the fermion fields, which are all flowed at the flow time $t$. Diagrams that give contribution proportional to the external momentum and thus vanish in our calculation are dropped. The external quark one-loop self-energy diagrams are not listed here, they are shown in figure.~\ref{fig:quarkself}}
\label{fig:sqpdf1}
\end{figure}
In this appendix, we consider the hadronic matrix element constructed using the operator defined in eq.~(\ref{eq:defqoperator}),
\begin{eqnarray}
\label{eq:quarkhmte}
h(z, P^z)= \langle P| \bar{\psi}(zv)\Gamma W(zv,0)\psi(0)| P\rangle,
\end{eqnarray}
where $P$ is the hadronic state with momentum $P^z$, $\Gamma = \gamma\cdot v$ or $\Gamma= \gamma^{\alpha}_{\perp}$.
The one-loop matching calculation for $h(z, P^z)$ from the gradient-flow scheme to the $\overline{\rm MS}$-like schemes was done in ref.~\cite{Monahan:2017hpu} with $P^z=0$ and full flow time dependence. 

However, the matching coefficient given by eq.~(44) of ref.~\cite{Monahan:2017hpu} does not lead to $z$ independent result in the small flow-time limit, 
which contradicts with our argument that we only need to do matching calculations for the local current operators stemming from the Wilson-line operators based on the one-dimensional auxiliary-field formalism.
Therefore, we re-calculate the matching coefficient in great detail and compare with the calculation done in ref.~\cite{Monahan:2017hpu}\footnote{Our results correct and supersede the results of ref.~\cite{Monahan:2017hpu}, for which an erratum is in preparation.}. 

The matching relation for $h(z, P^z)$ can be expressed as
\begin{eqnarray}
h^{\rm R}(z, t) = \mathcal{C}_q(t, z, \mu)h^{\overline{\rm MS}}(z, \mu),
\end{eqnarray}
 where we have dropped the exponentiated ``mass” correction $\delta m$ in the matching relation and choose $P^z =0$ for simplicity and comparison. 
 
 The Feynman diagrams that contribute to the  one-loop corrections for $h(z, P^z)$ in gradient-flow formalism are shown in figure.~\ref{fig:quarkself} and  figure.~\ref{fig:sqpdf1}.
The contribution coming from the quark self-energy diagrams in figure.~\ref{fig:quarkself} can be summarized as  
\begin{eqnarray}
\label{eq:appquarkself}
\mathcal{M}_{\mathring{\chi}}^{\rm R} =  \frac{\alpha_s}{4\pi}C_F\left[\frac{1}{\epsilon_{\rm IR}}+\log\left(2t\mu^2\right) + \gamma_E - \log(432) \right],
\end{eqnarray}
where the subscript $\mathring{\chi}$ indicates that we have used the ringed fermion field to remove the UV divergences.
\subsection{Wilson line self-energy diagram $(f_1)$}
 The one-loop Wilson line self energy diagram is shown as diagram $(f_1)$ in figure.~\ref{fig:sqpdf1} and its amplitude is given by
 \begin{eqnarray}
\mathcal{M}_{(f_1)}&=&-g^2\tilde{\mu}^{2\epsilon} C_F \int_0^z d\, s_1\int_0^{s_1} d\, s_2\, \int \frac{d^d k}{(2\pi)^d} \frac{1}{k^2}e^{-2t k^2} e^{i(k\cdot v)(s_2-s_1)},
\end{eqnarray}
which is finite at $d=4$. Therefore, we set $d=4$ and obtain,
\begin{eqnarray}
\label{eq:Wilsonselffull}
\mathcal{M}_{(f_1)}&=&-g^2C_F \int_0^z d\, s_1\int_0^{s_1} d\, s_2\,\int_0^{+\infty} dx \frac{1}{(2\pi)^4}e^{-\frac{(s_2-s_1)^2}{4(2t +x)}}\left(\frac{\pi}{2t +x}\right)^{2}\nonumber\\
&=&\frac{\alpha_s}{4\pi}C_F\times 2\left[\log\left(\bar{z}^2\right) +\gamma_E+2\left(1-e^{-\bar{z}^2}\right)-\text{Ei}\left(-\bar{z}^2\right)-2\sqrt{\pi}\bar{z}\, \text{erf}\left(\bar{z}\right)\right],
\end{eqnarray}
where $\bar{z} =  \frac{z}{r_F}$ with $r_F =\sqrt{8t}$ representing the flow radius, $\text{Ei}\left(x\right)$ is the exponential integral
\begin{eqnarray}
\text{Ei}\left(x\right) = -\int_{-x}^{+\infty} \frac{e^{-t}}{t}dt,
\end{eqnarray}
and $\text{erf}\left(\bar{z}\right)$ is the error function
\begin{eqnarray}
\text{erf}\left(\bar{z}\right) = \frac{2}{\sqrt{\pi}}\int_0^{\bar{z}} e^{-t^2} dt.
\end{eqnarray}
In the small flow-time limit, we have
\begin{eqnarray}
\label{eq:selfsmallflowlimit}
\mathcal{M}_{(f_1), t\rightarrow 0} = \frac{\alpha_s}{4\pi}C_F\times 2\left[\log\left(\bar{z}^2\right) +\gamma_E + 2-2\sqrt{\pi}\bar{z}\right],
\end{eqnarray}
where the last term is consistent with the one-lop result of $\delta m$ we have obtained in eq.~(\ref{eq:deltamresult}).
Notice that the $\overline{\rm MS}$ renormalized result of the Wilson line self energy at one-loop is given by \cite{Izubuchi:2018srq}
\begin{eqnarray}
\mathcal{M}_{(f_1)}^{\overline{\rm MS}} = \frac{\alpha_s}{4\pi}C_F\times 2 \left[2 + 2\gamma_E+ \log\left(\frac{\mu^2z^2}{4}\right)\right].
\end{eqnarray}
Dropping the linearly divergent term from $\mathcal{M}_{(f_1), t\rightarrow 0}$ and comparing with $\mathcal{M}_{(f_1)}^{\overline{\rm MS}} $, gives the contribution from diagram $(f_1)$ for the matching from the gradient-flow scheme to the $\overline{\rm MS}$ scheme in the small flow-time limit,
\begin{eqnarray}
\label{eq:WilsonselfGF2MS}
\mathcal{C}_{(f_1),\, t \rightarrow 0} =  \frac{\alpha_sC_F}{4\pi}\left[-2\log\left(2\mu^2te^{\gamma_E}\right)\right],
\end{eqnarray}
which is consistent with the result $\zeta_{h_v}$ we have obtained eq.~(\ref{eq:Zhvresult}).

\subsection{Vertex diagram $(f_2)$}
The amplitude of diagram $(f_2)$ is given by (including its mirror diagram),
\begin{eqnarray}
\mathcal{M}_{(f_2)} =2igC_F \int_0^z d\, s\int \frac{d^d k}{(2\pi)^d} \frac{-i\slashed{k}}{(k^2)^2} \left(ig\gamma\cdot v\right)\, e^{-2t k^2} e^{-i(k\cdot v)s}.
\end{eqnarray}
Since the transverse components of $\slashed{k}$ lead to vanishing momentum integration, we can replace $\slashed{k}$ with $(k\cdot v)(\gamma\cdot v)$ in $\mathcal{M}_{(f_2)}$, therefore, we have
\begin{eqnarray}
\label{eq:resultsdiagramd}
\mathcal{M}_{(f_2)} &=& 2g^2C_F \int \frac{d^d k}{(2\pi)^d} \frac{1}{(k^2)^2} \left(1-e^{-i(k\cdot v) z}\right)\, e^{-2t k^2} \nonumber\\
&=& \frac{\alpha_s}{4\pi}C_F\times 2\left[\log\left(\bar{z}^2\right) + \gamma_E - 1 -\text{Ei}\left(-\bar{z}^2\right) + \frac{1}{\bar{z}^2}\left(1-e^{-\bar{z}^2}\right)\right].
\end{eqnarray}
In the small flow-time limit, we have
\begin{eqnarray}
\mathcal{M}_{(f_2), t\rightarrow 0} &=&\frac{\alpha_s}{4\pi}C_F\times 2\left[\log\left(\frac{z^2}{8t}\right) + \gamma_E - 1 \right].
\end{eqnarray}
Comparing the corresponding result in the $\overline{\rm MS}$ scheme \cite{Izubuchi:2018srq},
\begin{eqnarray}
\mathcal{M}_{(f_2)}^{\overline{\rm MS}} &=&\frac{\alpha_s}{4\pi}C_F\times 2\left[2\gamma_E  +\log\left(\frac{\mu^2z^2}{4}\right) \right],
\end{eqnarray}
we obtain the contribution from diagram $(f_2)$ for the matching from the gradient-flow scheme to the $\overline{\rm MS}$ scheme in the small flow-time limit,
\begin{eqnarray}
\label{eq:quarkvertexGF2MS}
\mathcal{C}_{(f_2), t\rightarrow 0} = \frac{\alpha_s}{4\pi}C_F\left[-2\log\left(2\mu^2te^{\gamma_E}\right)  -2 \right],
\end{eqnarray}
which is consistent with the result of $\zeta_{\psi h_v}$ we have obtained in eq.~(\ref{eq:Zqhvresults}).

\subsection{Diagram $(f_3)$}
The amplitude of diagram $(f_3)$ can be expressed as,
\begin{eqnarray}
\mathcal{M}_{(f_3)} &=&(ig)^2\tilde{\mu}^{2\epsilon} C_F \int \frac{d^d k}{(2\pi)^d} \frac{\delta^{\mu\nu}\gamma^\mu(-i\slashed{k})\gamma^\alpha(-i\slashed{k})\gamma^\nu}{(k^2)^3}e^{-2t k^2}e^{-i(k\cdot v)z}\nonumber\\
&=& g^2C_F \left(d-2\right)\tilde{\mu}^{2\epsilon} \int \frac{d^d k}{(2\pi)^d} \frac{k^2\gamma^\alpha -2k^\alpha\slashed{k}}{(k^2)^3}e^{-2t k^2}e^{-i(k\cdot v)z}.
\end{eqnarray}
If $\alpha$ is parallel to $v$, dropping the tree-level factor $v^{\alpha}\gamma\cdot v $, we have
\begin{eqnarray}
\label{eq:resultsdiagrama0}
\mathcal{M}_{(f_3),\, \|}
&=& g^2C_F \left(d-2\right)\tilde{\mu}^{2\epsilon}  \int \frac{d^d k}{(2\pi)^d} \frac{\left[k^2-2(k\cdot v)^2\right]}{(k^2)^3}e^{-2t k^2}e^{-i(k\cdot v)z}\nonumber\\
&=&\frac{\alpha_s}{4\pi}C_F\bigg[-\frac{1}{\epsilon_{\rm IR}} - \log\left(\frac{\mu^2z^2}{4}\right) -2\gamma_E +3 \nonumber\\
&&\, \, \, \, \, \, \, \,\, \, \, \, \, \, \, \, \,  \, \, \,  +\text{Ei}(-\bar{z}^2) +\frac{3}{\bar{z}^4}\left(1-e^{-\bar{z}^2}\right) - \frac{1}{\bar{z}^2}\left(4-e^{-\bar{z}^2}\right)\bigg].
\end{eqnarray}
If $\alpha$ is transverse to $v$,  dropping the tree-level factor $\gamma^{\alpha}_{\perp}$, we have
\begin{eqnarray}
\label{eq:resultsdiagramaalpha}
\mathcal{M}_{(f_3),\, \perp}&=& g^2C_F \frac{d-2}{d-1}\tilde{\mu}^{2\epsilon}  \int \frac{d^d k}{(2\pi)^d} \frac{\left(d-3\right)k^2+2(k\cdot v)^2}{(k^2)^3}e^{-2t k^2}e^{-i(k\cdot v)z}\nonumber\\
&=&\frac{\alpha_s}{4\pi}C_F\bigg[-\frac{1}{\epsilon_{\rm IR}} - \log\left(\frac{\mu^2z^2}{4}\right) -2\gamma_E +1 \nonumber\\
&&\, \, \, \, \, \, \, \,\, \, \, \, \, \, \, \, \,  \, \, \,  +\text{Ei}(-\bar{z}^2) - \frac{1}{\bar{z}^4}\left(1-e^{-\bar{z}^2}\right) + \frac{1}{\bar{z}^2}
e^{-\bar{z}^2}\bigg].
\end{eqnarray}
In the small flow-time limit, we have
\begin{eqnarray}
\mathcal{M}_{(f_3),\, \|}^{t\rightarrow 0} &= &\frac{\alpha_s}{4\pi}C_F\bigg[-\frac{1}{\epsilon_{\rm IR}} - \log\left(\frac{\mu^2z^2}{4}\right) -2\gamma_E +3 \bigg],\\
\mathcal{M}_{(f_3),\, \perp}^{t\rightarrow 0} &= &\frac{\alpha_s}{4\pi}C_F\bigg[-\frac{1}{\epsilon_{\rm IR}} - \log\left(\frac{\mu^2z^2}{4}\right) -2\gamma_E +1 \bigg].
\end{eqnarray}
Noticing the corresponding $\overline{\rm MS}$ results are given by \cite{Izubuchi:2018srq},
\begin{eqnarray}
\mathcal{M}_{(f_3), \|}^{\overline{\rm MS}} &= &\frac{\alpha_s}{4\pi}C_F\bigg[-\frac{1}{\epsilon_{\rm IR}} - \log\left(\frac{\mu^2z^2}{4}\right) -2\gamma_E +3 \bigg],\\
\mathcal{M}_{(f_3), \perp}^{\overline{\rm MS}} &= &\frac{\alpha_s}{4\pi}C_F\bigg[-\frac{1}{\epsilon_{\rm IR}} - \log\left(\frac{\mu^2z^2}{4}\right) -2\gamma_E +1 \bigg],
\end{eqnarray}
which are exactly the same with the gradient flow results in the small flow-time limit. This indicates that diagram $(f_3)$ does not contribute to the matching from the gradient-flow scheme to 
the $\overline{\rm MS}$ scheme in the small flow-time limit, which supports our argument that we only need the matching for the local current operators stemming from the Wilson-line operators based on the one-dimensional auxiliary-field formalism.

\subsection{Final results}
Adding up the full result of each diagram in figure.~\ref{fig:sqpdf1} and the result of quark self-energy diagram in figure.~\ref{fig:quarkself}, which are given by eq.~(\ref{eq:Wilsonselffull}), eq.~(\ref{eq:resultsdiagramd}), 
eq.~(\ref{eq:resultsdiagrama0}), eq.~(\ref{eq:resultsdiagramaalpha}) and eq.~(\ref{eq:appquarkself}), dropping the linearly divergent term through ``mass” renormalization, we obtain the renormalized full result in the gradient-flow scheme
\begin{eqnarray}
\label{eq:quarkquasiGFfull}
\mathcal{M}_{q} = \frac{\alpha_s}{4\pi}C_F\left[a_{\Gamma}  + 3\log\left(\bar{z}^2e^{\gamma_E}\right)-\log(432)-4e^{-\bar{z}^2} - 3\text{Ei}(-\bar{z}^2) -4\sqrt{\pi}\bar{z}\text{erf}(\bar{z}) + 4\sqrt{\pi}\bar{z} \right],\nonumber\\
\end{eqnarray}
where
\begin{eqnarray}
a_{\gamma\cdot v}&=& 5 +\frac{3}{\bar{z}^4}\left(1-e^{-\bar{z}^2}\right) -\frac{1}{\bar{z}^2}\left(2+e^{-\bar{z}^2}\right),\\
a_{\gamma^{\alpha}_\perp}&=& 3 - \frac{1}{\bar{z}^4}\left(1-e^{-\bar{z}^2}\right) + \frac{1}{\bar{z}^2}\left(2-e^{-\bar{z}^2}\right).
\end{eqnarray}
The $\overline{\rm MS}$ renormalized  total result is 
\begin{eqnarray}
\mathcal{M}_{q}^{\overline{\rm MS}} = \frac{\alpha_s}{4\pi}C_F\left[a^\prime_{\Gamma}+ 3\log\left(\frac{z^2\mu^2}{4}\right)+6\gamma_E\right],
\end{eqnarray}
with $a^\prime_{\gamma\cdot v} = 7,\, a^\prime_{\gamma^{\alpha}_\perp} = 5$.

Then, we have the full matching coefficient, 
\begin{eqnarray}
\mathcal{C}_q (t, \mu, z)& =& 1 +\frac{\alpha_s}{4\pi}C_F\bigg[a_\Gamma  - a^\prime_\Gamma - 3\log\left(2\mu^2te^{\gamma_E}\right)-\log(432) - 4e^{-\bar{z}^2}\nonumber\\
&&\, \, \, \, \, \, \, \, \, \, \, \, \, \, \, \, \, \, \, \, \, \, \, \,\, \, \, \, \, \, - 3 \text{Ei}(-\bar{z}^2) -4\sqrt{\pi}\bar{z}\text{erf}(\bar{z}) + 4\sqrt{\pi}\bar{z}\bigg] +O(\alpha_s^2),
\end{eqnarray}
which gives the small flow-time limit,
\begin{eqnarray}
\mathcal{C}_{q,\, t\rightarrow 0}(t, \mu) = 1 -  \frac{\alpha_s}{4\pi}C_F\left[3\log\left(2\mu^2te^{\gamma_E}\right)+2+\log(432)\right],
\end{eqnarray}
which is indeed independent of $z$ and $\Gamma$.
It is straightforward to check that $\mathcal{C}_{q,\, t\rightarrow 0}(t, \mu) = c_{\psi h_v}^2$ as expected (see eq.~(\ref{eq:cpsihvfinal})).
\begin{figure}
\centering
\includegraphics[width =0.72\textwidth]{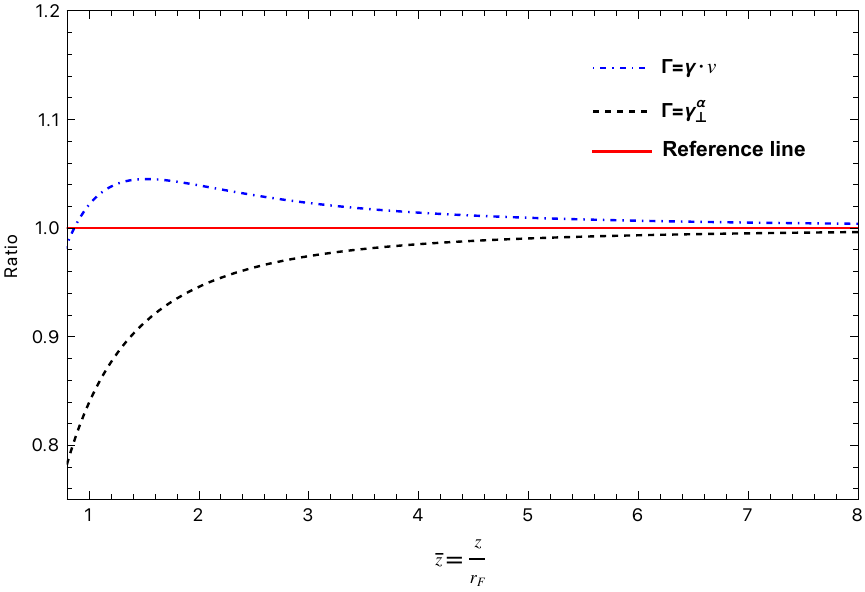}
\caption{The ratio of the one-loop correction result with full flow time dependence and the one-loop correction result in the small flow-time limit: ${\rm Ratio} =\frac{\mathcal{C}_q (t, \mu, z)-1}{\mathcal{C}_{q,\, t\rightarrow 0}(t, \mu) -1}$. The renormalization scale $\mu$ is chosen so that $\log\left(2t\mu^2e^{\gamma_E}\right)=0$.}
\label{fig:plot}
\end{figure}

To see how large the finite flow time effect will be, we plot the ratio of the one-loop corrections with full flow time dependence and the one-loop corrections in the small flow-time limit in figure.~\ref{fig:plot}.
As it is shown in figure.~\ref{fig:plot}, the finite flow time effect is very small as long as the flow radius $r_F = \sqrt{8t}$ is smaller than the distance $z$ ($\bar{z}>1$), which is usually satisfied in lattice gradient flow computations.

\bibliography{wgf.bbl}
\bibliographystyle{JHEP}

\end{document}